\documentclass[twocolumn,showpacs,amsmath,amssymb]{revtex4}
\usepackage{graphicx}
\usepackage{slashed}
\usepackage{pifont}
\usepackage{color}
\newcommand{\E}[1]{\times 10^{#1}}
\newcommand{\req}[1]{Eq.\,(\ref{#1})}
\newcommand{\beqn}{\begin{equation}}
\newcommand{\eeqn}{\end{equation}}
\newcommand{\beqnar}{\begin{equation}\begin{array}{r c l}}
\newcommand{\eeqnar}{\end{array}\end{equation}}
\newcommand{\eps}{\epsilon}
\newcommand{\veps}{\varepsilon}
\newcommand{\tr}{\mathrm{tr}\,}

\newcommand{\defn}{\mathrel{\mathop:}=}
\def\lambbar{{\mathchar'26\mkern-9mu\lambda}}

\newcommand{\Lag}{\ensuremath{\mathcal{L}}}
\newcommand{\Leff}{\ensuremath{V_{\mathrm{eff}}}}

\newcommand{\Tmm}{\ensuremath{\mathcal{T}}}
\newcommand{\Tmn}{\ensuremath{T_{\mu\nu}}}
\newcommand{\Tumn}{\ensuremath{T^{\mu\nu}}}

\newcommand{\gmn}{\ensuremath{g_{\mu\nu}}}
\newcommand{\gumn}{\ensuremath{g^{\mu\nu}}}
\newcommand{\condd}{\ensuremath{\langle  :\psi \bar \psi: \rangle}}
\newcommand{\cond}{\ensuremath{\langle \bar\psi  \psi \rangle}}
\newcommand{\condphi}{\ensuremath{\langle   \phi \phi^*  \rangle}}
\newcommand{\Imag}{\mathtt{Im}}

\newcommand{\gBI}{\varepsilon_{\mathrm{BI}}}
\newcommand{\calS}{\mathcal{S}}
\newcommand{\calP}{\mathcal{P}}
\newcommand{\csch}{\mathrm{csch}\:}
\newcommand{\Fumn}{F^{\mu\nu}}
\newcommand{\Humn}{H^{\mu\nu}}
\newcommand{\Fmn}{F_{\mu\nu}}
\newcommand{\dFumn}{\widetilde{F}^{\mu\nu}}
\begin{document}
\title{Dark Energy Content of Nonlinear Electrodynamics}
\author{Lance Labun and Johann Rafelski}
\affiliation{Department of Physics, University of Arizona, Tucson, Arizona, 85721 USA, 
 and\\ Department f\"ur Physik der Ludwig-Maximillians-Universit\"at M\"unchen und\\
  Maier-Leibniz-Laboratory, Am Coulombwall 1, 85748 Garching, Germany}
\date{29 August 2009}
\begin{abstract}
Quasi-constant external fields in nonlinear electromagnetism generate a 
contribution to the energy-momentum tensor with the form of dark energy.  
To provide a thorough understanding of the origin and strength of the effects,
we undertake a complete theoretical and numerical study of the energy-momentum
tensor $T^{\mu\nu}$ for nonlinear electromagnetism.  
The Euler-Heisenberg nonlinearity due to quantum fluctuations of spinor 
and scalar matter fields is considered and contrasted with the properties 
of classical nonlinear Born-Infeld electromagnetism.  We also address 
modifications of charged particle kinematics by strong background fields.
\end{abstract}

\maketitle
%
\section{Introduction}\label{sec:INT}
Recent analysis has constrained the
dark energy to have almost exactly the characteristics of a 
cosmological constant: equation of state $w \equiv p/\rho \approxeq -1$ and
spatially homogeneous distribution~\cite{Komatsu:2008hk,Serra}.  
This means that the dark energy is present in the energy-momentum tensor 
proportional to $\gumn$  as has often 
been discussed in the context of vacuum energy~\cite{Weinberg:1988cp}.
Proportionality to $\gumn$ does not exclude that dark energy originates 
in properties of ponderable fields and matter; a nonvanishing
trace of the energy-momentum tensor $T_{\mu}^{\mu}=\gumn\Tmn$, where 
\begin{equation}\label{Teff}
\Tmn  =
  \frac{2}{\sqrt{-g}} \frac{\delta}{\delta \gumn}\int d^4x \sqrt{ -g} \:\Leff,
\end{equation}
signals presence of a dark energy term in \Tmn.
 
A good candidate for dark energy is a physical theory that, without 
interactions, has a traceless energy-momentum tensor.  Extensions of this 
theory including interactions can then 
be analyzed for the generation of an energy-momentum trace, and by comparison
with the bare theory, the physics of the energy-momentum 
trace is more easily extracted. An effort in this direction 
has been made in the context of vacuum structure 
of quantum chromodynamics~\cite{Schutzhold:2002pr}, we show here
that standard quantum electrodynamics (QED) has similar features 
which are more easily accessible. 

Our point of departure is thus Maxwell electromagnetism, 
whose energy momentum tensor,
\beqn\label{classicalTmn}
\Tmn^{\rm Max} = 
  \gmn\frac{1}{4}F_{\alpha\beta}F^{\alpha\beta}
	-F_{\mu\lambda}F_{\nu}^{\phantom{\nu}\lambda}
\eeqn
is explicitly traceless.  Even in absence of external sources the
Maxwell field equations are incomplete  due to interaction with the 
electron-positron vacuum fluctuations which are present at the 
length scale $ \lambbar_e \equiv \hbar/m_ec$. 
At distances of comparable magnitude ($\lambda \simeq \lambbar_e$) 
these are vacuum polarization 
effects which impact precision atomic physics experiments.  For 
long distance  $(\lambda \gg \lambbar_e )$ one  obtains the 
nonlinear effective theory of the photon studied in depth by 
Euler, Kockel and Heisenberg, and Schwinger~\cite{EKH,Schwinger51}.

This nonlinear Euler-Heisenberg (EH)  theory of electromagnetism is 
just one of many possible effective actions.
Beyond the EH-QED framework, we can imagine writing down a more 
complete theory containing all effective interactions, reducing in the long 
wavelength limit at the classical level to Maxwell's equations. 
Born-Infeld (BI) theory, designed to regulate
point-particle-induced divergences~\cite{BI}, can be thought of as an effort 
to provide a more complete theory of electromagnetism, though it too is now 
considered as an effective theory~\cite{SBI} arising from string theory. 

We show how in nonlinear electromagnetism, the classical energy-momentum tensor
\req{classicalTmn} is modified by two quantities: a dielectric function $\veps$,
which scales the contribution of the Maxwell energy-momentum in the total, 
and the trace of energy momentum tensor $T^{\mu}_{\mu}$, which is a 
new contribution.  We have proposed that $T^{\mu}_{\mu}$  arising in QED 
can in this regard be viewed as the modification of the vacuum energy by the 
presence of electromagnetic fields~\cite{Labun:2008gm},
and indeed the connection of dark energy with a form of vacuum energy has 
been discussed before~\cite{Weinberg:1988cp,Schutzhold:2002pr}. In this work
we fully develop this conjecture, evaluate the dielectric function and compare 
the behavior of EH and BI  nonlinear theories.

As is well known, a theory with a traceless energy-momentum
tensor is invariant under scale changes.  Maxwell's electromagnetism 
is the prime example: the energy-momentum tensor is traceless and
indeed the classical theory of radiation is scale invariant.  In a gauge 
theory, the related
conformal symmetry can be spontaneously broken, the study of which
has a long and distinguished history, originating with the rise of
quantum chromodynamics (QCD) as the theory of strong 
interactions~\cite{Chanowitz:1972da,Crewther:1971bt}.   
We will rederive here some results of the 
extensions of these studies to QED~\cite{Adler77,Collins77}. 

Scale invariance can be broken by the explicit appearance of a 
dimensionful quantity in the theory, such as the mass of the electron in QED.
In fact, any nonlinear electromagnetism requires a 
scale with the dimension of an electrical field $eE_0$, 
in order to render the Lagrangian dimensionally consistent.
We express this scale in terms of a mass $M$:
$$ eE_0 \equiv \frac{(M c^2)^2}{\hbar c}$$
For the Euler-Heisenberg (EH) effective action the introduction of $M$ is 
a natural step to take as the nonlinearity is 
of quantum origin and $M\simeq  m/\sqrt{\alpha}$, where $\alpha=1/137$ is the
usual fine structure constant. For the BI theory it is a matter of convenience 
to use mass rather than length as the scale, converting one into another 
using $\hbar$.

If indeed BI theory is a weak-field limit of string theory in which
the nonlinearity is a consequence of high mass quantum fluctuations~\cite{SBI}
the appearance of $\hbar$ would be appropriate, and the associated scale could
be as large as the Planck mass, $M_{\rm Pl}=1.2\E{19}$\,GeV. It should be noted
that current experimental limits as well as EH nonlinearity probes  a scale 
below 100 MeV, thus a  string related BI theory maybe quite removed from 
the present experimental reality. We compare EH and BI  theories mainly because
their behavior is very different. 

In QED and BI, the presence of scale which breaks the conformal symmetry  
is explicit, and  the energy-momentum trace is not ``anomalous,'' unlike 
the case of QCD.  The scale of the nonlinearity is, as we shall show, 
the determining factor of  $T^{\mu}_{\mu}$.  This fact is the simple yet
important and original theoretical observation presented in this paper. 
Having thus suggested the inter-connection of dark energy, conformal 
symmetry and the presence of scale in the theory, we leave issues specific 
to conformal symmetry to future work and here focus on the physics of 
$T_{\mu}^{\mu}$ and its origins in nonlinearity of the theory. 

We derive in Section~\ref{sec:Tmn} the field energy-momentum tensor 
and explicitly connect its trace to nonlinearity of the electromagnetic theory.
To compare relative magnitudes and suggest new constraints on a 
Born-Infeld-type completion of electromagnetism, in Section~\ref{sec:BI} 
we evaluate the BI modifications to the Maxwell energy-momentum tensor.  
In clarification of conflicting statements present in the literature, we 
begin our discussion of quantum electrodynamics in Section~\ref{sec:QED} with 
a new derivation of the relationship between the electron-positron condensate
and the energy-momentum trace based on the explicit origin of the trace in 
nonlinearity of the theory.  Extending a technique of resummation of the 
action introduced by M\"uller, et al.~\cite{Mueller77}, we then provide 
complete numerical evaluations of the condensate, energy-momentum trace and 
dielectric function for QED and spin-0 quantum electrodynamics.  These 
evaluations display striking analytical features not before apparent in the  
Euler-Heisenberg functions.  We compare BI and QED contributions to
$T^{\mu}_{\mu}$ and show that QED vacuum fluctuations remain dominant given 
the experimental constraints.
 
In the final Section \ref{sec:Tmm-kin} of this report, we discuss 
the kinematics of charged particles moving in external fields.  
The vacuum of a nonlinear theory is studied
as a ponderous medium with nonlinear response.  
The Lorentz force is preserved, but the breaking of the superposition 
principle results in effective potentials for charged particles moving in 
external fields that are not automatically obtained from the Lorentz force.

\section{Energy-Momentum Tensor of Nonlinear Electromagnetism}
\label{sec:Tmn}
%
Setting from now on $\hbar =c=1$, we can consider the effects of 
$E_0\defn M^2/e$ mass $M$ or length $l=1/M$ scale, 
where $M$ can be as large as a string theory scale or as small as the mass 
of the electron. The consequences are best seen writing the
effective action in the form 
\begin{equation}\label{Veff}
\Leff \equiv -\mathcal{S} + 
M^4\:\overline{f_{\rm eff}}\!\left(\frac{\calS}{M^4},\frac{\calP}{M^4}\right) 
   \underset{M \to \infty}{\longrightarrow} -\mathcal{S}
\end{equation}
presented here as a function of the Lorentz scalar and pseudoscalar
\begin{subequations}
\begin{eqnarray}
\mathcal{S} &\defn&
 \frac{1}{4}F_{\kappa\lambda}F^{\kappa\lambda}=\frac{1}{2}(B^2-E^2); \\
\mathcal{P} &\defn& \frac{1}{4}F^*_{\kappa\lambda}F^{\kappa\lambda}=E\cdot B.
\end{eqnarray}
\end{subequations}
As noted, classical, linear electromagnetism must constitute the limit
of \req{Veff} for fields small as measured in units of $E_0$, and
only the classical theory does not require a dimensioned scale.

\subsection{Dielectric function and Trace}
\label{ssec:eps+tr}
To understand the implications of the dimensioned scale we consider the 
explicit form of the energy momentum tensor~\eqref{Teff}, 
separating the traceless Maxwell part.  For a general
function $\Leff(\calS,\calP)$, we obtain 
\begin{eqnarray} \label{SchTmn}
\Tmn &=&
   \left(-\frac{\partial \Leff}{\partial \mathcal{S}}\right)
   (g_{\mu\nu} \mathcal{S}- F_{\mu\lambda}F_{\nu}^{\phantom{\lambda}\lambda})
 \\[0.15cm]
& &\quad-\gmn \left(\Leff-\mathcal{S}\frac{\partial \Leff}{\partial\mathcal{S}}
     -\mathcal{P}\frac{\partial \Leff}{\partial \mathcal{P}} \right). \nonumber
\end{eqnarray}
Comparison with \req{classicalTmn} shows the
energy-momentum tensor of Maxwell's electromagnetism is modified by a 
dielectric function, $-\partial\Leff/\partial\calS$, to be discussed below.
Using Eq.\,(\ref{Veff}) we simplify the second term to
\beqn \label{dVdm}
\left(\Leff-\mathcal{S}\frac{\partial \Leff}{\partial \mathcal{S}}
           -\mathcal{P}\frac{\partial \Leff}{\partial \mathcal{P}}\right)=
\frac{1}{4} M\frac{\partial \overline{f_{\rm eff}}}{\partial M} ,
\eeqn 
This form is very useful because it provides a simple means of calculating 
the trace directly from the effective action.  
The importance of \req{dVdm} lies in its distillation of the {physical}
source of conformal symmetry breaking in any nonlinear theory of
electromagnetism: 

Terms linear in the invariant $\mathcal{S}$ cannot contribute to the right 
side of \req{dVdm} since they cancel explicitly on the left side of \req{dVdm}.
Such contributions must therefore be omitted from $\Leff$ in the study of 
energy-momentum trace, and hence we have introduced the barred 
$\overline{f_{\rm eff}}$ to denote the nonlinear components of the 
effective potential.  Letting $\Leff^{(1)}$ denote the remaining linear 
terms in the Lagrangian, we have the decomposition
\beqn \label{L-decomp}
\Leff = \Leff^{(1)} + M^4 \overline{f_{\rm eff}}.
\eeqn
The lowest power in $\cal S, P$ is 2 due to 
preservation of both parity and charge conjugation symmetry, which 
respectively require an action even under parity transformations and even in 
the coupling $e$, hence even in the field strengths (already true of any
Lorentz scalar).  A nonzero imaginary part of the action entails 
breaking of time reversal symmetry, though the trace \Tmm will be small under
laboratory conditions.  These symmetry arguments imply that for field
strengths below the critical scale $E \ll m^2$, the 
energy-momentum trace must be at least 4th order in the fields.
 
Solving the partial differential equation
\beqn\label{tracelessPDE}
\Leff-\mathcal{S}\frac{\partial \Leff}{\partial \mathcal{S}}
           -\mathcal{P}\frac{\partial \Leff}{\partial \mathcal{P}}=0
\eeqn	   
displays one obvious class of nonlinear Lagrangians that have traceless
energy-momentum tensors, namely
\beqn\label{notr-theories}
\Leff = \calS \sum_{n=-\infty}^{+\infty}a_n\left(\frac{\calS}{\calP}\right)^{n}
\eeqn
Comparison with \req{Veff} reveals the reason: such Lagrangians are conformal.
Since this class is non-perturbative in at least one of the field invariants 
and we are interested in having an energy-momentum trace, we exclude 
theories of the form in \req{notr-theories} as well any other non-perturbative
actions that may satisfy \req{tracelessPDE}, despite their inherent 
interest. 

In view of Eqs.\,(\ref{SchTmn},\ref{dVdm}), we summarize
\begin{subequations} \label{SchTmn1}
\begin{eqnarray}
&&\Tmn = \veps\Tmn^{\rm Max} +\gmn \frac{1}{4}\Tmm \\
&&\veps \equiv -\frac{\partial \Leff}{\partial \mathcal{S}},\qquad  \label{Tan}
 T_{\mu}^{\mu}\equiv \Tmm = -M\frac{d\overline{f_{\rm eff}}}{dM}
\end{eqnarray}
\end{subequations}
Here $\veps$ is the dielectric function, $\Tmn^{\rm Max}$ the
Maxwell energy-momentum tensor, and $\Tmm$ the energy-momentum trace.
Interestingly, $\Tmm/4 \leftrightarrow \lambda/2$ provides a dark energy
or Einstein-like cosmological constant, while the traceless part is the same 
as in Maxwell theory, up to the multiplicative dielectric function. 

Eq.\,(\ref{Tan}) though derived here for the case of electromagnetism has 
a much wider domain of validity. The implications of the separation in 
Eq.\,(\ref{SchTmn1}) have been previously noted in context of the photon 
propagation effects~\cite{Dittrich98,Shore96}, but $\Tmm$was not given in the 
form Eq.\,(\ref{Tan}), nor has in any form the trace $\Tmm$ been computed.
We shall below demonstrate how the concomitant identities Eq.\,(\ref{Tan})
provide new physical insight in understanding the quantum effects and 
their connection to nonlinearity of the action.

The alternative and equivalent  representation, often used in the
electromagnetism of nonlinear media 
(for example see section 8 of~\cite{BialynickiBirula:1975yp}),
\beqn\label{alt-Tmn}
T^{\mu\nu} = H^{\mu\lambda}F^{\nu}_{\phantom{\nu}\lambda}-g^{\mu\nu}\Lag
\eeqn
is using the displacement tensor
\beqn\label{displacement}
\Humn = \frac{\partial \Lag}{\partial \Fmn} = 
  \frac{\partial\Lag}{\partial\mathcal{S}}\Fumn
	  +\frac{\partial\Lag}{\partial\mathcal{P}}\dFumn .
\eeqn
The trace is now distributed into several components
\beqn\label{alt-Tmm}
\Tmm = H^{\mu\nu}F_{\mu\nu}-4\Lag 
     = \vec E\cdot \vec D + \vec B\cdot \vec H - 4\Lag.
\eeqn
The origins and properties  of $\Tmm$ are 
obscured by the constitutive relation $\Humn(\Fumn,\dFumn)$.
Our expression \req{SchTmn1} evidences the departure from the
classical theory more clearly in the context we consider.

\subsection{Stress-Energy Density}
\label{sec:SE}

Some further notable properties of the stress-energy density of the 
nonlinear EM field will be collected here.  
The energy density in the frame of the metric, 
i.e. the quantity entering Einstein's equations, is 
$$T^{00}= \frac{\veps}{2}(E^2+B^2)+\frac{1}{4}\Tmm.$$
The EM-stresses $T^{ij}$ have the same structure as in the Maxwell theory, 
as is already evident from the format of Eq.\,(\ref{BI-uf2}):
\beqn\label{stress}
T^{ij}= \veps T^{ij}_{\rm Max}-\delta_{ij}\frac{1}{4}\Tmm
\eeqn
The trace $\Tmm$ acts to compensate the forces 
$T^{ij}_{\rm Max}$ tearing the field sources apart in Maxwell
electromagnetism. For this reason, for example in BI theory the $T^{ij}$ 
vanishes allowing a stable charged particle without material stresses.  
A sufficient condition for this to be true is that the point particle 
solution satisfies $ \lim_{r\to 0}r^3T^{00}=0$~\cite{Rafelski:1972fi}.

Because the energy momentum tensor is conserved 
we have $T_{\mu\nu,\nu}\equiv \partial \Tmn/\partial x^{\nu}=0$, which is 
a covariant relation true in any frame.  A differential conservation law 
leads to an integral conservation law by integration over the observer's
hypersurface:
\beqn\label{uf0}
\int_1^2 d^4x\frac{\partial \Tumn}{\partial x^{\nu}}=0,\:\:\mathrm{or}\:\:
 \int_1 d^3\sigma_\nu\Tumn = \int_2 d^3\sigma_\nu\Tumn,
\eeqn
i.e. the energy-momentum flow through surface 1 is the same as later 
through surface 2.  It is common to choose an observer at rest in laboratory
so that $ d^3\sigma_{\nu} =u_{\nu} d^3x $, 
with $u^{\nu}=(1,0,0,0)$. 

For nearly homogeneous fields we can
omit the 3-volume and consider a conserved 4-momentum density of the EM field  
$$
p^{\mu}_{\rm Max} = 
     u_{\nu}\Tumn_{\rm Max} \to \left( (E^2+B^2)/2, \vec E\times \vec B \right)
$$
finding the well known result for the rest energy and Poynting vector of the
classical field.  This result is easily generalized to the nonlinear
electromagnetism:
\beqn\label{BI-uf2}
p^{\mu} =
\left(\veps \frac{E^2+B^2}{2}+\frac{\Tmm}{4}, \veps \vec E\times \vec B \right),
\eeqn
showing the appropriateness of calling $\veps$ the dielectric 
function, since it plays the role of the dielectric
constant $\vec E=\veps\vec D$ when considering electric charge in vacuum.
For Maxwell electromagnetism  $\veps =1$ and $\Tmm=0$.

It is an elementary exercise to show that in the Maxwell limit the proper 
energy density, or its ``mass density,'' is 
\begin{subequations}\begin{eqnarray}
p_{\mu}^{\rm Max}p^{\mu}_{\rm Max}
    \!&\!=\!&\! \frac{(E^2+B^2)^2}{4} -( \vec E\times \vec B)^2 \\
    \!&\!=\!&\! \frac{(E^2-B^2)^2}{4}+(\vec E\cdot \vec B)^2
    \!=\!\calS^2+\calP^2.
\end{eqnarray}\end{subequations}
Generalizing to nonlinear EM theory, we find the local mass
density of the field to be
\begin{eqnarray} 
\label{uf1}
u_f\equiv \sqrt{p_\mu p^\mu} &=& \sqrt{(\calS^2+\calP^2)\veps^2+(\Tmm/4)^2}
\end{eqnarray}
$\Tmm/4 \leftrightarrow \lambda/2$ provides thus both a 
`dark energy' and a `mass density' of the electromagnetic field.


\begin{figure*}
\hspace*{-.5cm} \includegraphics[width=0.5\textwidth]{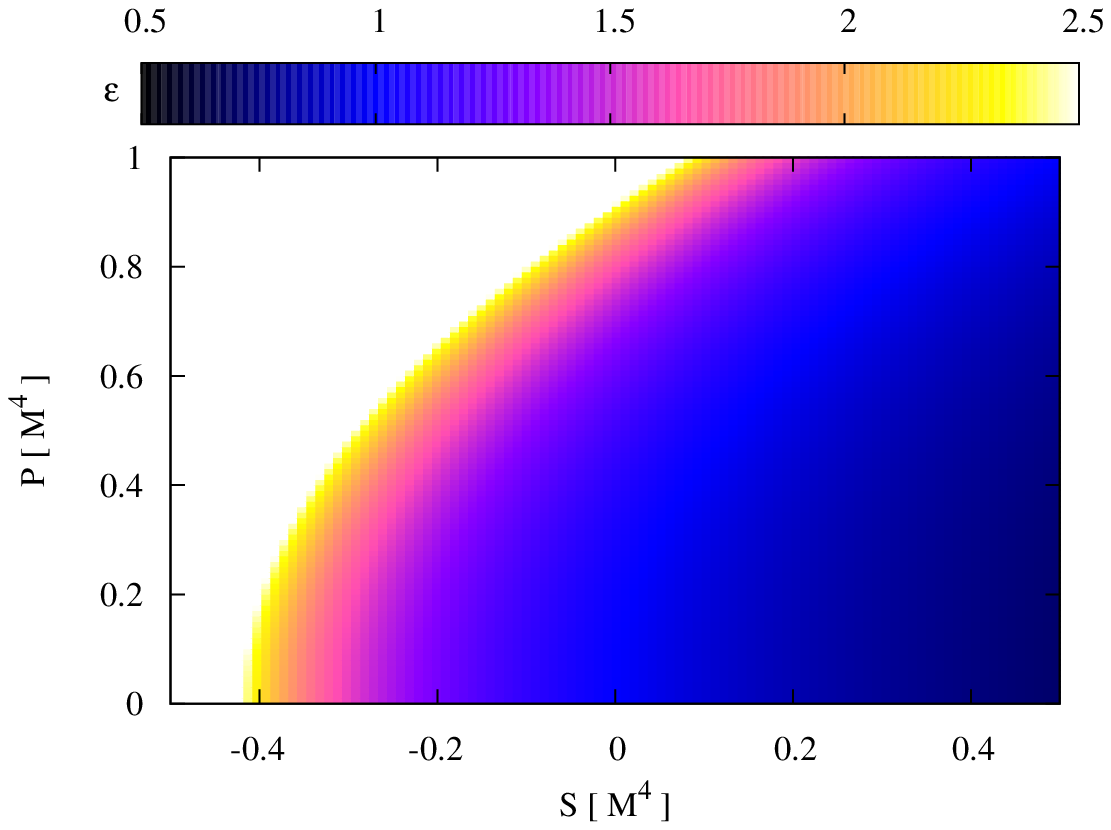}
 \includegraphics[width=0.48\textwidth]{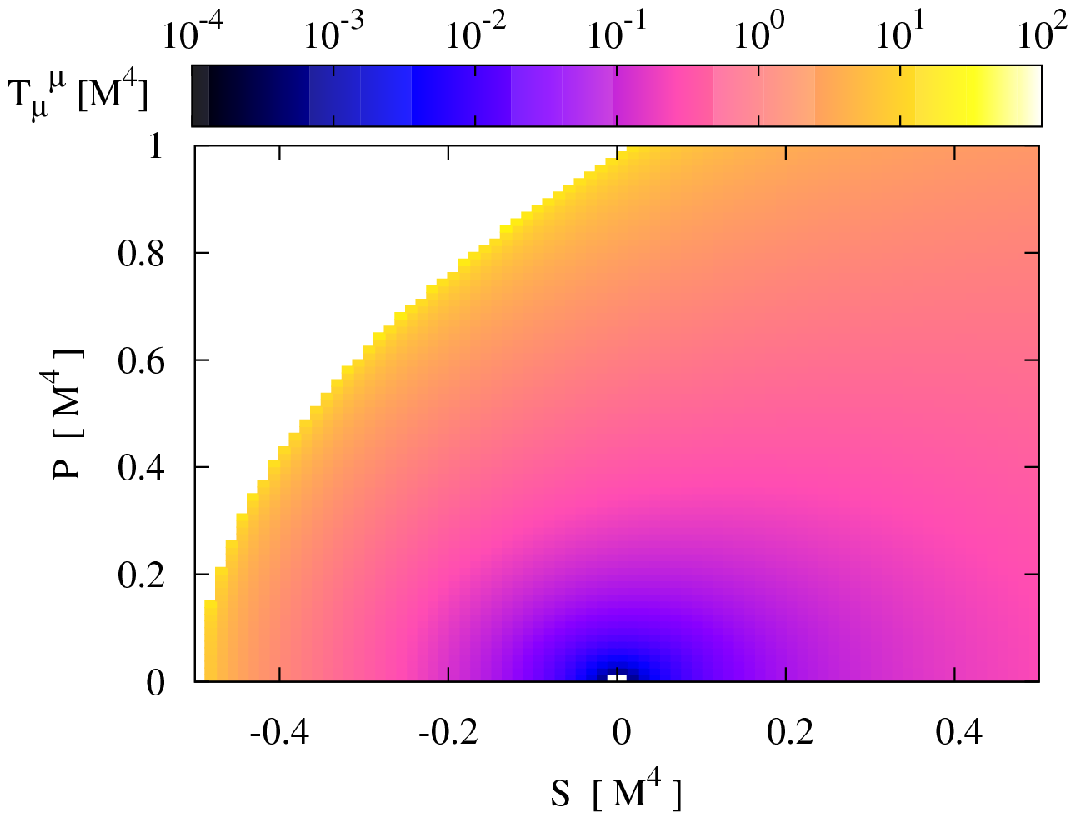}
\caption{The departure of the Born-Infeld energy-momentum tensor from 
that of the Maxwell $T^{\mu\nu}_{\rm Max}$: At left the dielectric function,
\req{gBI} and at right the trace, \req{BI-Tmm}.  
Field strength invariants and energy densities are in units of $M^4$. }
\label{fig:BI}       
\end{figure*}

\section{Born-Infeld Electromagnetism}
\label{sec:BI}
As demonstrated in the preceding discussion, an intrinsically nonlinear 
theory of electromagnetism (in most cases) entails an energy-momentum trace, 
and we begin by studying a non-quantum example 
of a nonlinear alternative to Maxwellian theory.
Historically, Born-Infeld electromagnetism was introduced in order to solve 
the infinite self-energy (and self-stress) problem of a point-like electron
arising in consideration of the radial electric field $E_r$ of a point 
charge $q$: 
$$
U=\int d^3x \frac{1}{2}E_r^2\to \infty,\quad {\rm for}\quad E_r=\frac{q}{r^2}
$$
To remedy this, Born and Infeld took inspiration from special relativity,
considering the action (note that we follow the modern 
convention, opposite in sign to the original paper~\cite{BI}, and also, 
recall remarks about the scale $M$ above Eq.(\ref{Veff})): 
\begin{eqnarray} \label{BI-Lag}
\Leff^{\rm (BI)}&=&M^4\left( 1-\sqrt{ 1+2\calS/M^4-(\calP/M^4)^2} \right)\\  
\label{BI-LagG} &=&M^4\left(\sqrt{-g}-\sqrt{-h} \right), \\[0.2cm] 
 h&=&\det h_{\mu\nu},  \quad    h_{\mu\nu} =g_{\mu\nu} +\frac{F_{\mu\nu}}{M^2},
\end{eqnarray}
where the particular combination of $S,P$ terms derives from the extension 
of the space-time metric with the antisymmetric field tensor, $F_{\mu\nu}$.  
In the weak field (infinite mass) limit Maxwell's theory indeed 
arises,\ $V^{\rm (BI)}\to -\calS$.  

For the Born-Infeld case the dielectric function is 
\beqn \label{gBI}
 \gBI=-\frac{\partial \Leff^{\rm (BI)}}{\partial S}   
        = [ 1+2\calS/M^4 - (\calP/M^4)^2 ]^{-1/2},
\eeqn
which exhibits a formal analogy to the $\gamma$-factor familiar
from special relativity, though with two different limits as
$\calS$ or $\calP$ respectively approach the limiting value $M^4$ 
(see figure~\ref{fig:BI}, left).  The dielectric function goes over from
suppression ($\veps<1$) to augmentation ($\veps > 1$), 
when the magnetic component of the field becomes subdominant, which 
corresponds to crossing the line $2\calS = \calP^2$ from the lower right.

As presented, \req{BI-Lag} is in the form required by \req{Veff}. 
However, the BI action 
contains no terms linear in $\mathcal{S}$; $\Tmm$ is identically
$-M(d\Leff/dM)$.
We obtain for the BI energy-momentum trace using the relation \req{dVdm}
\begin{eqnarray}\label{BI-Tmm}
\Tmm^{(\mathrm{BI})} &=& 4M^4(\gBI(1+\calS/M^4 1)-1),\\[0.4cm]\notag
&=& 4M^4\left(\sqrt{\frac{1+2\mathcal{S}/M^4+
    (\mathcal{S}/M^4)^2}{1+2\mathcal{S}/M^4-(\mathcal{P}/M^4)^2}}-1\right)
\end{eqnarray}
which in the latter form is manifestly positive-definite, just like the
cosmological constant.
For small fields we expand the second form in Eq.\,(\ref{BI-Tmm}) to obtain 
\beqn \label{BIlim}
\Tmm^{(\mathrm{BI})} \to 
	\frac{2}{M^4}\frac{\calS^2+\calP^2}{\sqrt{ 1+2\calS/M^4  }}
\eeqn
In figure~\ref{fig:BI} we show the dielectric function and 
energy-momentum trace for strengths up to the maximum field strength.
The functional behavior of both is smooth, though
a $\calS, \calP$ functional asymmetry develops at large
values of the fields.  

In contrast to the classical theory analyzed here, we note a recent report
suggesting that the quantized BI theory as studied on the lattice may be
conformally symmetric~\cite{Kogut:2006ur}.

\section{Euler-Heisenberg Electromagnetism}
\label{sec:QED}
The  Euler-Heisenberg effective action is well known:
\beqn\label{SpinLag}
\hspace*{-0.2cm}\Leff^{\rm f} = \!\int_{0+\delta}^{\infty}  \!\!\frac{ds\,e^{-m^2s}}{8\pi^2s^3}
\left(1-eas \cot(eas)\:ebs \coth(ebs) \right)
\eeqn
for Dirac fermions, and
\beqn \label{ScalLag}
\Leff^{\rm s} = \!\int_{0+\delta}^{\infty} \!\!
\frac{ds\,e^{-m^2s}}{16\pi^2s^3}(eas \csc(eas)\:ebs\:\mathrm{csch}\:(ebs)-1)
\eeqn
for charged scalars, in which 
\beqn a\defn \sqrt{\sqrt{\calS^2+\calP^2}-\calS},\:\:\mathrm{and}\:\: 
	b\defn \sqrt{\sqrt{\calS^2+\calP^2}+\calS}.
\eeqn
The characteristic strength of fluctuations in the 
matter field is made explicit  in the appearance
of $m$ which is the mass of the matter particle which 
has been integrated out, in QED it is the mass of the electron. 

For vanishing $\calP$ we have $a\to \! \sqrt{|\calS|-\calS}$ and 
$b\to \! \sqrt{|\calS|+\calS}$.  Thus when $E=0$ we find $a\to 0, b\to |B|$ 
and when $B=0$ we find $a\to |E|, b\to 0$.  
In this sense, $b$ plays the role of the 
generalized (Lorentz invariant) magnetic field and $a$ that of the generalized 
electric field.  The constant subtraction $\pm 1$ removes the (divergent) zero
point energy of free electrons and positrons. The difference in normalization
reflects the doubling of the number of degrees of freedom for spin-1/2
particles, and the overall sign corresponds to the difference in sign of 
vacuum fluctuations between bosons and fermions.

\subsection{The  condensate $\cond$ and trace $\Tmm$}
\label{ssec:cond}
Equation\:\eqref{Tan} provides a direct means of calculating the 
energy-momentum trace, but its connection to vacuum structure in QED and
its deformation by the applied fields which induce the EH nonlinearity is  
encoded in $m(d\Leff/dm)$. Consider the Feynman 
boundary condition Green's function of the fluctuating matter field 
in presence of the electromagnetic field, which determines the vacuum fluctuations,
\beqn\label{dLdm-prop}
m\frac{\partial\Leff(x,m)}{\partial m} = 
 im\lim_{\substack{\epsilon\to 0}} \tr [S_F(x+\epsilon,x-\epsilon,m)-S_F^{(0)}]
\eeqn
Here $\epsilon$ is a time-like vector, and $S_F^{(0)}$ is the free field 
Feynman Green's function.  In general:
\beqn \label{propdef}
S_F(x,x')  = -i\langle T(\psi(x')\bar \psi(x)) \rangle. 
\eeqn

We rewrite the right side of \req{dLdm-prop} using the elementary form of 
Wick's decomposition theorem 
$$ T(\psi(x')\bar \psi(x)) =:\psi(x')\bar \psi(x): +\langle 0| T(\psi(x')\bar \psi(x))|0 \rangle,$$  
where the normal ordering is with respect to the `no-field' vacuum.  Taking the
expectation value of this relation at a single space-time point in the
`with-field' vacuum $| \rangle$ we find 
\begin{eqnarray} \label{cond1}
\condd \hspace*{-0.15cm}&=&\hspace*{-0.15cm}
  \langle  |T(\psi(x)\bar \psi(x))|  \rangle -\langle 0| T(\psi(x)\bar \psi(x))|0 \rangle \\
&=& iS_F(x,x) -iS_F^0(x,x) \notag 
\end{eqnarray}
with the same $\epsilon$-limit as in \req{dLdm-prop} implied for equal
propagator arguments.  Usually the trace is implied by commuting the fields 
and the normal ordering symbols omitted, 
$$m\condd \to -m\cond ,$$
hiding the important operational definition~\req{cond1} of what we now 
recognize as the fermionic condensate.  Thus, we see that the condensate 
derives from the difference of normal ordering in the no-field (also called
perturbative) vacuum and the with-field vacuum.  Furthermore, one must 
keep in mind that the subtraction of the unperturbed vacuum term follows 
directly from application of the rules of QED and is not a consequence of 
arbitrary removal of zero-point energy.  For this reason our discussion
has no bearing on the zero-point energy of quantum field theory and/or its 
gravitational coupling.  

Equations\:\eqref{propdef}, \:\eqref{dLdm-prop} and \:\eqref{cond1} then 
combine with the result 
\beqn \label{dLdm}
m\frac{d\Leff}{dm} =-m\cond,
\eeqn
a known and widely used relation in non-perturbative QCD. We will evaluate 
it for the case of electrons in presence of external electromagnetic fields.

We undertook the derivation of the Fermi condensate in the preceding
section in order to emphasize that the condensate is not in general
the energy-momentum trace, though there is a relationship. 
Using Eqs\:\eqref{L-decomp}, \:\eqref{Tan}, and \:\eqref{dLdm} we find 
\beqn\label{Tmm-decomp}
 \Tmm =m\frac{d\Leff^{(1)}}{dm} + m\cond.
\eeqn 
If and only if $\Leff^{(1)}$ is  just the action of  
classical electromagnetism,  the first 
term on the right hand side, linear in $\calS$,  vanishes.

However,  the coupling
to a fluctuating quantum field complicates the issue significantly
in that it produces a contribution linear in $\calS$ with a 
coefficient which is a function of $m$ and renormalization scale. It is 
the logarithmic divergence of Eqs.\:\eqref{SpinLag} and \:\eqref{ScalLag}
which in the limit $\delta\to 0$ is also proportional to $\mathcal{S}$ and 
thus appears in $\Leff^{(1)}$.  The standard procedure is to absorb the 
divergence into the definition of the charge in the process of charge
renormalization.  Eqs.\:\eqref{SpinLag} and \eqref{ScalLag}
assume a cutoff regulator  which of course also  introduces a scale. 
Alternatively one can use dimensional 
regularization, the spinor effective potential can be written
\beqn\label{SpinLag-dimreg}
\hspace*{-0.1cm}\Leff^{\rm f}  = \!\int_{0}^{\infty}  
     \!\!\frac{ds\,e^{-m^2s}}{8\pi^2s^{3-\epsilon}}
\left(1-eas \cot(eas)\:ebs \coth(ebs) \right) 
\eeqn
and similar for the charged scalar case.

For any finite $\epsilon$, \req{SpinLag-dimreg} is finite and can be
differentiated with respect to $m$, and hence the quantity 
$m(d\Leff^{\rm f,s}/dm)$ is finite, allowing the consideration of a 
vanishing $\epsilon$.  In particular, the differentiation $m(d\Leff/dm)$ 
renders the lowest order contribution $m(d\Leff^{(1)}/dm)$
finite.  The condensate and energy-momentum trace  
are thereby independent of renormalization procedure (as they should be),
and obey the nontrivial relationship expressed by \req{Tmm-decomp}.  
The finiteness of $\Tmm$ for fermions is also discussed at length by 
Adler et al~\cite{Adler77}.  

Separation of the term linear in $\calS$ in the Fermi case shows
\beqn \label{Maxwell-corr}
m\frac{d\Leff^{(1)}}{dm} = 
-\frac{e^2}{8\pi^2}\frac{b^2-a^2}{3}\int_0^{\infty} \frac{ds}{s}e^{-m^2s}
= \frac{2\alpha}{3\pi}\cal{S}.
\eeqn
Thus for the EH effective action, the 
decomposition in Eq.\:\eqref{L-decomp} becomes
\beqn\label{linearCor2} 
m\frac{d\Leff^{\rm f}}{dm}
  =  m\frac{d\overline{\Leff^{\rm f}}}{dm} +\frac{2\alpha}{3\pi} \mathcal{S}
\eeqn
and in turn with emphasis on properties of the vacuum:
\beqn\label{QED-Tmm}
 \Tmm^{\rm f} = \frac{2\alpha}{3\pi}\langle \mathcal{S}\rangle + m\cond.
\eeqn
Equation\;(\ref{QED-Tmm}) corresponds to Eq.(2.17) in \cite{Adler77}. It is 
of importance to note that there is considerable cancellation between 
the two terms on the right hand side~\cite{Schutzhold:2002pr}.

The essential relation \req{QED-Tmm} must be preserved at any number of 
loops in the effective action in its suitable generalization.  In particular, 
if the condensate were evaluated to two loops, the coefficient of the first 
term must become the two-loop $\beta$-function, i.e.
\beqn \label{beta-gen}
\frac{2\alpha}{3\pi} \rightarrow
 \beta(\alpha)=\frac{2\alpha}{3\pi}+\frac{\alpha^2}{2\pi^2}+...
\eeqn

Our result \req{QED-Tmm} is not obvious if one evaluates 
the energy-momentum trace after renormalization has been carried out.
The logarithmically divergent term 
in $\Leff$ is
\beqn\label{linear} 
\Leff^{(1)} =
-\frac{e^2}{8\pi^2} \frac{b^2-a^2}{3}\int_{0+\delta}^{\infty} \!\frac{ds}{s}e^{-m^2s}
=\frac{\alpha}{3\pi}\mathcal{S}\ln (m^2\delta)
\eeqn
in which $\delta=1/M^2$, some large mass or momentum extraneous to QED.  
Before presenting the EH action, 
this term is  absorbed in the process of charge renormalization.  To 
restore its contribution one must realize the mass dependence of charge 
renormalization. The relation of \req{linear} to the QED $\beta$ 
function~\cite{Adler77} demonstrates why use of the renormalized $\Leff$
with the incorrect identification $\Tmm= md\Leff/dm$ leads to the correct
result as shown in \req{QED-Tmm}, just as it was developed for
QCD~\cite{Chanowitz:1972da,Crewther:1971bt,Schutzhold:2002pr}.

\subsection{Properties of $\Tmm$ in QED}
\label{ssec:QED}

We obtain the explicit form of the condensate in external
fields combining \req{dLdm} and \req{SpinLag}
\begin{eqnarray}\label{cond2}
-m\cond =  m^2\!\int_{0}^{\infty} \frac{ds\,e^{-m^2s}}{4\pi^2s^2}
&&\hspace*{-0.2cm}\left(eas \cot(eas)\,\right.\times\\
&&
\:ebs  \coth(ebs)-1).\notag
 \end{eqnarray}
The condensate vanishes in the absence of field, as it should, since 
$x\cot x\to 1$ and $x \coth x \to 1$ for $x\to 0$. The   term  quadratic in 
the fields has been discussed above, it must be subtracted 
to arrive at the integral representation of
the energy-momentum trace:
\begin{eqnarray} \label{QED-Tmm-int}
\Tmm^{\rm f}  =
- \frac{m^2}{4\pi^2}\!\int_{0}^{\infty}\!\!\frac{ds\,e^{-m^2s}}{s^2}
&&\hspace*{-0.2cm}
\left(eas \cot(eas)   \,  \times  \right.  \\ 
&& \hspace*{-1.3cm}
ebs\coth(ebs)-1-\frac{e^2}{3}(b^2-a^2)s^2\left. \right).\notag 
\end{eqnarray}

To study the integrals \req{cond2} and \req{QED-Tmm-int}, we introduce 
a transformation that will be helpful in dealing with the non-analyticities 
generated by the electric field.  The detailed calculations are carried
out in the Appendix~\ref{app:stats}, and we simply state here their results.
The integral representations obtained along the way have better convergence
properties than \req{cond2} or \req{QED-Tmm-int}, particularly at strong 
fields $B,E \sim E_0$.

In a magnetic background field, the condensate can be written
\beqn \label{statkern-B}
-m\cond = 
-\frac{m^4}{2\pi^2\beta'}\int_0^{\infty}\frac{\ln(1-e^{-\beta' s})}{s^2+1} ds,
\eeqn
where $\beta':=\pi m^2/eB=\pi/(B/E_0)$, with $E_0=m^2/e$.
The energy-momentum trace is obtained by removal of the leading 
term quadratic in the field (recall \req{QED-Tmm}), with the result that
\beqn \label{Tmm-B-step2}
\Tmm^{\rm f} =
-\frac{m^4}{2\pi^2\beta'}\int_{0}^{\infty} \frac{s^2\ln(1-e^{-\beta' s})}{1+s^2}ds
\eeqn
is manifestly positive definite.

For the electric field, the poles are resummed into a logarithmic winding
point, with the result
\beqn \label{statkern-E}
-m\cond = \frac{m^4}{2\pi^2\beta}\int_{0}^{\infty} \frac{\ln(1-e^{-\beta s})}{1-s^2-i\eps}ds
\eeqn
where $\beta := \pi/(E/E_0)$.  
The trace in the pure electric background is
\beqn\label{Tmm-spinorE}
  \Tmm^{\rm f}  = \frac{m^4}{2\pi^2\beta}\int_0^{\infty}\!\!
            \frac{s^2\ln(1-e^{-\beta s})}{1-s^2-i\eps}ds.
\eeqn

\begin{figure*}
\includegraphics[width=.48\textwidth]{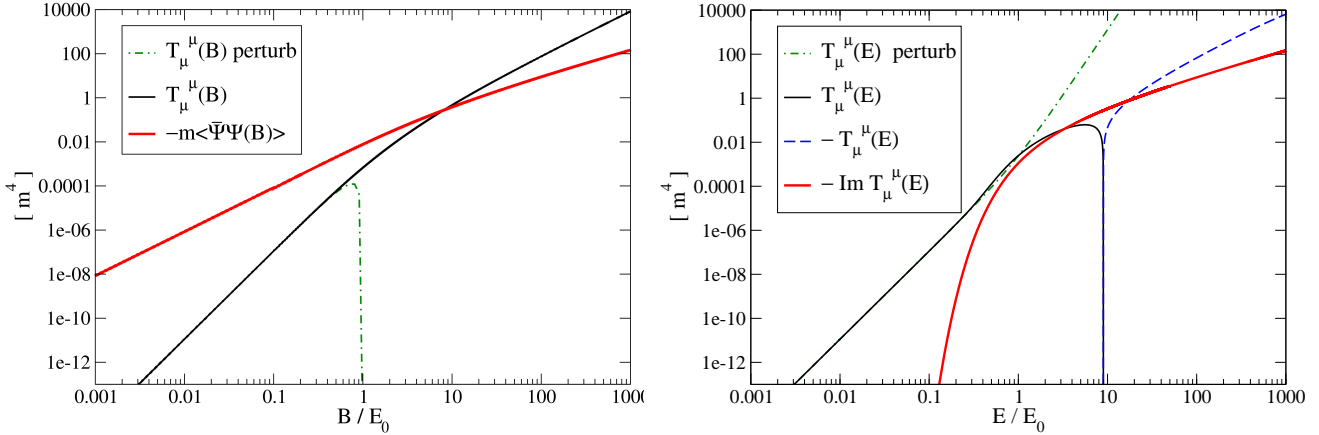}
\includegraphics[width=.48\textwidth]{labun-spin-E.eps}
\caption{The trace $T_{\mu}^{\mu}$ normalized by the electron mass
$m_e^{4}$ for spin-1/2 matter fields in magnetic (left) and electric (right)
fields, evaluated perturbatively (dash-dot line) from the first two terms in
\req{QED-Tmm-weakfield} and exactly (solid line) from 
Eqs.\,\eqref{Tmm-B-step2} and \eqref{Tmm-spinorE}.  At left,
the condensate in magnetic-only backgrounds is included for comparison
of magnitudes, and at right, the imaginary part present for electric fields
is included.
The trace changes sign from positive to negative at $E=9E_0$ and for higher
fields the negative ($-T^{\mu}_{\mu}$) is plotted.}
\label{spin-cond}       
\end{figure*}

Equations\:\eqref{statkern-E} and~\eqref{Tmm-spinorE} could equivalently 
be obtained by taking $B \rightarrow iE$ in the respective magnetic 
expressions.  The condensate behaves as $(eE)^2$ for small fields, but
the poles displayed in
Eq.\:\eqref{cond-E-step1} give the condensate and the energy-momentum trace
a nonzero imaginary part, which can be evaluated from \req{statkern-E} or
\eqref{Tmm-spinorE} recalling the identity
$$ \frac{1}{x-i\eps} = \mathrm{PV}\frac{1}{x}+i\pi \delta(x). $$
As befits its role contributing to the proper mass of the nonlinear 
electromagnetic field (recall \req{uf1}), 
\beqn \label{cond-imag}
\Imag\,\Tmm^{\rm f}
 = -m^2\frac{eE}{4\pi^2}\!\sum_{n=1}^{\infty}\!\frac{1}{n}e^{-\frac{n\pi E_0}{E}}
 = \frac{m^4}{4\pi\beta}\ln(1-e^{-\beta}),
\eeqn
is manifestly negative and strongly suppressed for field strengths 
less than $0.1E_0$.
This is consistent with direct differentiation of the positive  
imaginary part of the action $\Leff^{\rm f}$ evaluated by Schwinger
$$
\Imag\,\Leff^{\rm f} 
=\frac{(eE)^2}{8\pi^3}\sum_{n=1}^{\infty} \frac{1}{n^2}e^{-\frac{n\pi E_0}{E}}.
$$

Observe that $m\cond$ in external magnetic fields is negative, 
as is manifest in \req{statkern-B}, while the energy-momentum trace, 
\req{Tmm-B-step2} is positive.  With the perspective that the energy-momentum
trace represents the energy arising from deformation of the vacuum, a negative 
value would imply that the vacuum state is unstable, for example to the
spontaneous generation 
of strong magnetic fields if $\Tmm$ were identically $m\cond$.

The energy-momentum trace in general combinations of electric and magnetic
can also be cast \req{gen-QED-Tmm} reminiscent of our prior study of the 
special cases of electric and magnetic fields.
Using $b$ in the definition of $\beta'$ and $a$ in $\beta$, the numerically
useful representation is:
\begin{subequations}\label{gen-QED-Tmm2}
\begin{eqnarray} 
&&\Tmm^{(QED)}=\frac{m^4}{2\pi^2}(I_a+I_b) \\
I_a &=& 
\int_0^{\infty}\!\!\frac{ds\,s^2}{1-s^2-i\epsilon} \sum_{k=1}^{\infty} 
    \frac{e^{-sk\beta }}{k\beta} \frac{\pi k\beta}{\beta'}\coth{\frac{\pi k \beta}{\beta'}} , \\
I_b &=& 
\int_0^{\infty}\!\! \frac{ds\,s^2}{1+s^2} \sum_{k=1}^{\infty} 
    \frac{e^{-sk\beta'}}{k\beta'} \frac{\pi k\beta'}{\beta}\coth{\frac{\pi k \beta'}{\beta}}. 
\end{eqnarray}
\end{subequations}
A further resummation (see appendix~\ref{app:stats}) displays a more 
statistical form of the integrands:
\begin{subequations}\label{gen-QED-Tmm3}
\begin{eqnarray}
I_a &=& -b\sum_{n,\sigma}\int_0^{\infty} \frac{s+\frac{1}{2}\ln\left(\frac{1-x+i\eps}{1+x-i\eps}\right)}{e^{\beta\mathcal{H}_{\sigma}}-1}\,ds \\
I_b &=& a\sum_{n,\sigma}\int_0^{\infty} \frac{s-\mathrm{Arctan}\,s}{e^{\beta'\mathcal{H}_{\sigma}'}-1}\,ds
\end{eqnarray}
\end{subequations}
where $\sigma=\pm 1$ and $n=0,1,2...$ in accordance with the Landau
levels apparent in the quasi-Hamiltonians
\begin{subequations}\label{f-Hamiltonians}
\begin{eqnarray}
m^2\mathcal{H}_{\sigma} &=& m^2s+(2n+1-\sigma)eb \\
m^2\mathcal{H}_{\sigma}' &=& m^2s+(2n+1-\sigma)ea.
\end{eqnarray}
\end{subequations}
The slower convergence of $I_a$ ($I_b$) for $b/a \ll 1$ ($a/b \ll 1$) means
\req{gen-QED-Tmm3} is not as advantageous as \req{gen-QED-Tmm2} for 
numerical work.
These expressions clarify the electric and magnetic field limits and thereby 
the correspondence between figures\:\ref{spin-cond} and\:\ref{fig:gen-Tmm} 
below when one also recalls  $z\coth z\to 1$ for $z\to 0$.

\subsection{Numerical Evaluation of QED \Tmm}\label{ssec:QEDnums}
\label{ssec:Nums}

We evaluate numerically the condensate and the energy-momentum trace for 
arbitrary constant, homogeneous background electromagnetic field.  We consider 
first a magnetic only background field in Eq.\:\eqref{cond2}, a form which is 
easily integrated numerically.  The behavior is displayed in 
figure~\ref{spin-cond} 
as the solid upper (red) line.  Noting that $f(x)=(x \coth x -1)/x^2$ is 
essentially constant for $x < 1$, we recover the quadratic dominance 
$-m\cond \propto (eB)^2$ for small fields.  Figure~\ref{spin-cond} confirms 
the small field (mass-dominant) quadratic behavior and large field 
(field-dominant) linear behavior~\cite{Shushpanov}.  The results of 
numerically integrating\:\eqref{statkern-E} also appear in 
figure~\ref{spin-cond}, but since the result is opposite in sign to the magnetic
field case, we show the negative of the result as the lower (black) curve.

Eqs.\:\eqref{Tmm-spinorE} and\:\eqref{Tmm-B-step2} are plotted on the right in 
figure~\ref{spin-cond}, and interestingly, we see that at $E=9E_0$ 
$\Tmm$ changes sign from positive at low fields to negative at high 
fields.  This result is not apparent in the perturbative expansion, but 
can be understood from the Cauchy-Riemann equations in view of the rapidly
changing imaginary part.  Having the appearance of a $\pi$ phase flip, the 
feature may be related to the rapid dissolution (attosecond timescale) 
of fields at magnitudes surpassing $E_0$~\cite{Labun08}.  The methods 
developed here are not suited to the dynamics implied by such fields and the 
processes used to obtain such field strengths.  For comparison, we also plot 
the weak field expansions, derived from the original EH expression with the 
power-law semi-convergent expansions of cot and coth.  The first two terms 
for the condensate and energy-momentum trace are
\beqn \label{QED-cond-weakfield}
-m\cond \approx
   \frac{m^4}{12\pi^2}\frac{\mathcal{E}^2}{E_0^2}
      -\frac{m^4}{90\pi^2}\frac{\mathcal{E}^4}{E_0^4}+\ldots
\eeqn 
\beqn \label{QED-Tmm-weakfield}
\Tmm^{\rm f} \approx 
  \frac{m^4}{90\pi^2}\frac{\mathcal{E}^4}{E_0^4}
    -\frac{4m^4}{315\pi^2}\frac{\mathcal{E}^6}{E_0^6} +\ldots\:
\eeqn 
where $\mathcal{E}^2=B^2$ or $\mathcal{E}^2= -E^2$ for magnetic or electric
fields, respectively.

\begin{figure}
  \includegraphics[width=0.48\textwidth]{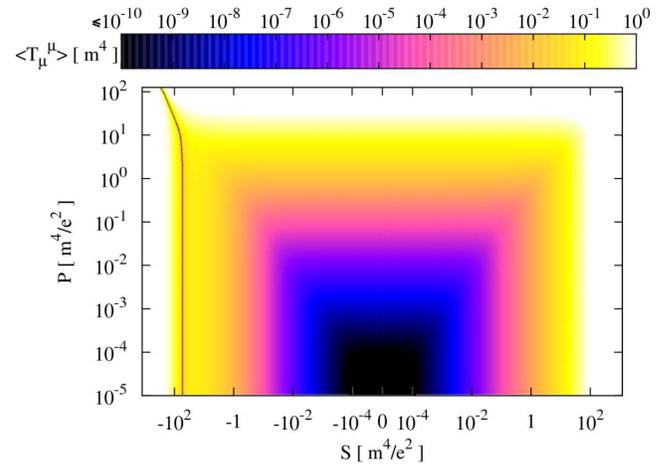}
\caption{The spinor energy-momentum trace for general E,B fields, 
parameterized by the Lorentz invariants, $\mathcal{S,P}$ as defined 
below Eq.\:\eqref{Veff}.  As only $\mathcal{P}^2$ appears in $\Leff$, we 
plot only positive $\mathcal{P}$.  Although the figure has a logarithmic 
scale, the trace crosses zero at the dark (grey) line, going from positive 
to negative for very large electric fields $\mathcal{S}\ll-1$, 
as seen also in fig. \ref{spin-cond}.  (Color online.)}
\label{fig:gen-Tmm}       
\end{figure} 
%

The condensate and energy-momentum trace for more general field configurations
are evaluated using the rapidly convergent sums in Eq.\:\eqref{gen-QED-Tmm}, 
and the results displayed in figure~\ref{fig:gen-Tmm}.  An examination of
the condensate for arbitrary field configurations, which does include the 
term linear in $\calS$ (but is not displayed here), shows nontrivial features
near the line $\calS=\calP$.  This is reflecting on the
expectation~\cite{Dunne02}, of the influence of zero modes, known to be 
present when the field is self-dual, i.e. $\mathcal{S=P}$.  Comparison of the
appearance of  figure~\ref{fig:gen-Tmm} with the BI result, 
figure~\ref{fig:BI} reveals a profound difference in these results, with 
the vacuum fluctuation effective action
 being more `edgy', and suggesting the possibility of a 
more singular behavior in the full multiloop strong field case.

The weak field expansions in general electromagnetic backgrounds give
\beqn \label{QED-cond-gen-weakfield}
-m\cond \approx
   \frac{m^4}{6\pi^2}\calS-\frac{m^4}{90\pi^2}(4\calS^2+7\calP^2)+\ldots
\eeqn 
\beqn \label{QED-Tmm-gen-weakfield}
\Tmm^{\rm f} \approx 
  \frac{m^4}{90\pi^2}(4\calS^2+7\calP^2)
   -\frac{4m^4}{315\pi^2}\calS(8\calS^2+13\calP^2) +\ldots\:
\eeqn
using $\calS, \calP$ normalized to the natural field strength, $E_0$.  We can
compare  perturbative EH result\:\eqref{QED-Tmm-gen-weakfield} with the Born-Infeld 
weakfield $\Tmm$ Eq.\:\eqref{BI-Tmm}, but because the coefficients of 
$\calS^2$ and $\calP^2$ do not match in QED, we obtain two values 
for the BI-type limiting mass, in obvious notation:
\beqn
\begin{array}{r c l}
M_{\calS}=&\sqrt[4]{\frac{45}{16\alpha^2}}m&=15.16m,\: \mathrm{or}\\[0.3cm]
M_{\calP}=&\sqrt[4]{\frac{45}{28\alpha^2}}m&=13.18m.
\end{array}
\eeqn

\subsection{$\Tmm$ of Scalar QED}
\label{ssec:scalQED}
\begin{figure*}
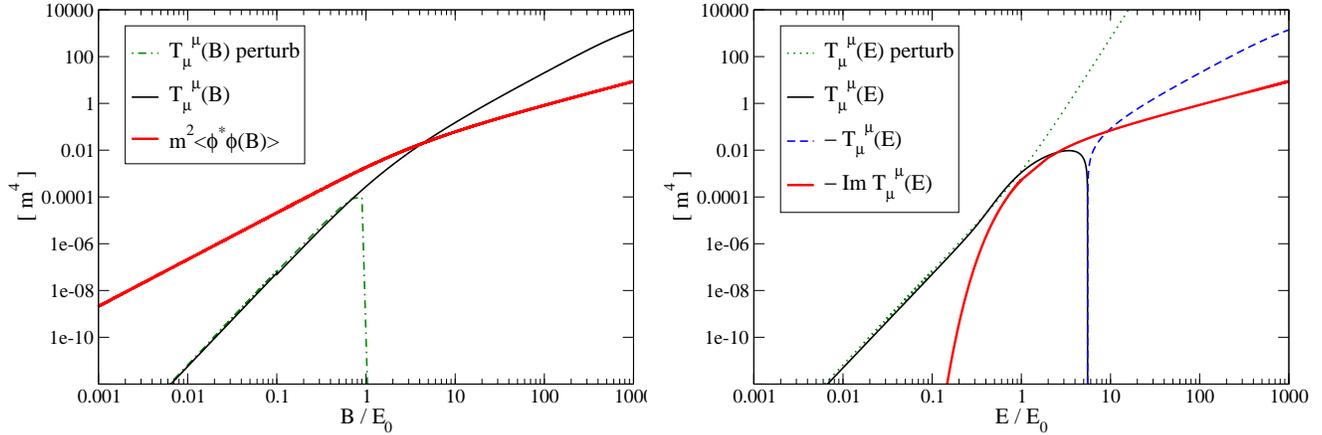

\includegraphics[width=.48\textwidth]{labun-scal-B.eps}
\includegraphics[width=0.48\textwidth]{labun-scal-E.eps}
\caption{For spinless particle fluctuations, we display the 
the energy-momentum trace in units of $m_e^{4}$ for both 
magnetic-only (left) and electric-only (right) backgrounds.  
Again, the condensate is displayed for comparison in the magnetic 
case, and the imaginary part of the trace present in the electric background.
Note the change in sign of the trace is now at $E=5.5E_0$.}
\label{scal-cond}       
\end{figure*}

The Euler-Heisenberg-Schwinger calculation of the effective action is 
easily extended to the case of a charged scalar field.  The spin-0 particle
has no Pauli spin coupling $\sigma_{\mu\nu}F^{\mu\nu}$ in the Hamiltonian, 
and the proper time integral is evaluated for the Klein-Gordon equation 
with covariant derivative,
\beqn\label{scalar-Hamiltonian}
(D^2+m^2)\phi = 0, \qquad
D_{\mu} \equiv \partial_{\mu}-ieA_{\mu},
\eeqn
leading to \req{ScalLag} displayed above.
The condensate takes the slightly different form, $m^2\condphi$.
Analogous manipulation of the proper time representations 
as above (see section\:\ref{ssec:cond}) leads to an identity similar
to\:\eqref{dLdm}:
\beqn \label{scal-dLdm}
m^2\condphi = m\frac{\partial\Leff^{\rm s}}{\partial m}.
\eeqn
In turn, we have
\beqn \label{scal-Tmm}
\Tmm^{\rm s} = \frac{\alpha}{6\pi}\calS -m^2\condphi,
\eeqn
which displays for the scalar field the large cancellation 
between the photon and matter condensates.
Then, since the explicit form of the energy-momentum 
tensor, Eq.\:\eqref{SchTmn}, remains valid, we need 
only manipulate the action of scalar electrodynamics.

The alternating sign in the meromorphic expansions of the $\csc$ and 
$\mathrm{csch}$ functions (see appendix~\ref{app:stats}) leads upon partial
integration to the opposite sign, 
or opposite ``statistics'' in the logarithm $\ln(1\pm e^{-\beta s})$, as was
already remarked upon by~\cite{Mueller77}.  Thus, 
we have for magnetic only background fields
\beqn \label{scal-condB}
m^2\condphi = 
 \frac{m^4}{4\pi^2\beta'}\int_0^{\infty}\frac{\ln(1+e^{-\beta' s})}{1+s^2}ds
\eeqn
and for electric only fields
\beqn \label{scal-condE}
m^2\condphi= 
\frac{m^4}{4\pi^2\beta}\int_0^{\infty}\frac{\ln(1+e^{-\beta s})}{1-s^2-i\eps}ds
\eeqn 
Results of numerically integrating equations\:\eqref{scal-condB}
and\:\eqref{scal-condE} appear in figure~\ref{scal-cond}.
As before, the condensate appears with opposite signs when comparing the 
electric and magnetic backgrounds, and the pole structure remains consistent 
for electric and magnetic fields, irrespective of particle type.  The 
imaginary part induced by the electric background similarly becomes
\beqn \label{scal-imag}
\Imag\,m\frac{d\Leff^{\rm s}}{dm} = 
m^2\frac{eE}{8\pi^2}\!\sum_{n=1}^{\infty}\! \frac{(-1)^n}{n}e^{-\frac{n\pi E_0}{E}} 
	= \frac{-m^4}{8\pi\beta}\ln(1+e^{-\beta}),
\eeqn
which is again negative in continued agreement with the role of $\Tmm$ 
in the proper mass-energy of the nonlinear electromagnetic field.

Finally, the energy-momentum trace generated by the scalar field quantum
fluctuations is
\beqn \label{Tmm-scalarB}
\Tmm^{\rm s}   = 
 \frac{m^4}{4\pi^2\beta'}\int_0^{\infty}\frac{s^2\ln(1+e^{-\beta' s})}{1+s^2}ds,
\eeqn
in the magnetic background, and
\beqn\label{Tmm-scalarE}
\Tmm^{\rm s}   = 
-\frac{m^4}{4\pi^2\beta}\int_0^{\infty}\frac{s^2\ln(1+e^{-\beta s})}{1-s^2-i\eps}ds
\eeqn
in the electric background.  Similarly positive for all 
but the highest electric fields, the scalar 
energy-momentum trace is exhibited in figure~\ref{scal-cond} on right.  
The effect of the fermionic 
statistics is apparent in the shift of the zero crossing to $E\simeq 20.2E_0$.
In the figure, we also compare the weak field expansions, 
\beqn \label{scalQED-cond-weakfield}
m^2\condphi \approx
   \frac{m^4}{48\pi^2}\frac{\mathcal{E}^2}{E_0^2}-
     \frac{7m^4}{1440\pi^2}\frac{\mathcal{E}^4}{E_0^4}+\ldots
\eeqn 
\beqn \label{scalQED-Tmm-weakfield}
\Tmm^{\rm s} \approx 
  \frac{7m^4}{1440\pi^2}\frac{\mathcal{E}^4}{E_0^4}-
  \frac{31m^4}{5040\pi^2}\frac{\mathcal{E}^6}{E_0^6} +\ldots\:
\eeqn 
where $\mathcal{E}^2=B^2$ or $\mathcal{E}^2= -E^2$ 
for magnetic or electric fields, respectively.

\begin{figure}
  \includegraphics[width=0.48\textwidth]{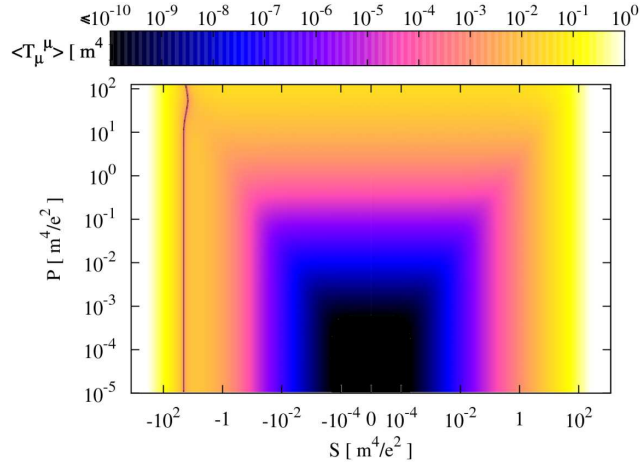}
\caption{The scalar energy-momentum trace for general E,B fields,
Eq.\:\eqref{gen-scalQED-Tmm}, parameterized by the Lorentz invariants.  
The dark (grey) vertical line indicates the change in sign of the trace to
negative values for very large electric fields. 
The transition is present up to arbitrary values of $\mathcal{P}$ 
as in fig.~\ref{fig:gen-Tmm}.  (Color online.)}
\label{fig:scal-genTmm}       
\end{figure} 
The results for general electric and magnetic backgrounds are 
\begin{subequations}\label{gen-scalQED-Tmm}
\begin{eqnarray}
\hspace*{-.8cm}
&&\Tmm^{\rm s}
  \!\! =\!\!-\frac{m^4}{4\pi^2} \left( I_a + I_b \right); \\[0.3cm]
\hspace*{-.8cm}
I_a\!\!
 &=&\!\!\! \int_0^{\infty}\!\!\!\frac{ds\,s^2}{1-s^2-i\epsilon} \sum_{k=1}^{\infty} \!
      \frac{(-1)^k e^{-sk\beta }}{k\beta}\frac{\pi k\beta}{\beta'}\csch{\frac{\pi k\beta}{\beta'}},\\[0.3cm]  
\hspace*{-.8cm}
I_b\!\! 
 &=&\!\!\! \int_0^{\infty}\!\!\! \frac{ds\,s^2}{1+s^2} \sum_{k=1}^{\infty} \!  
   	\frac{(-1)^k e^{-sk\beta'}}{k\beta'}\frac{\pi k\beta'}{\beta}\csch{\frac{\pi k\beta'}{\beta}}.  
\end{eqnarray}
\end{subequations}
which is plotted in figure~\ref{fig:scal-genTmm}.  
The statistical form analogous to \req{gen-QED-Tmm3} can be obtained from
that expression by changing to the fermionic $+$ sign in the denominator and
omitting the spin sum and $\sigma$ term in the quasi-Hamiltonians
\req{f-Hamiltonians}; the spectral functions 
$$
s+\frac{1}{2}\ln\left(\frac{1-x+i\eps}{1+x-i\eps}\right)~~\mathrm{and}~~
s-\mathrm{Arctan}\,s
$$ 
remain unchanged.
The weak field expansions 
for general electromagnetic backgrounds are
\beqn \label{scalQED-cond-gen-weakfield}
m^2\condphi \approx
   \frac{m^4}{24\pi^2}\calS-\frac{m^4}{360\pi^2}(7\calS^2+\calP^2)+\ldots
\eeqn 
\beqn \label{scalQED-Tmm-gen-weakfield}
\Tmm^{\rm s} \approx 
  \frac{m^4}{360\pi^2}(7\calS^2+\calP^2)
   -\frac{m^4}{630\pi^2}\calS(31\calS^2+11\calP^2)\ldots\:
\eeqn
using $\calS, \calP$ normalized to the natural field strength, $E_0$.  Again, 
we compare with the Born-Infeld weak-field energy-momentum trace 
(see Eq.\:\eqref{BI-Tmm}), and as above, the coefficients of $\calS^2$ and 
$\calP^2$ are not the same.  For the scalar quantum theory then, the two 
corresponding values for the BI-type limiting mass are
\beqn
\begin{array}{r c l}
M_{\calS}&=\sqrt[4]{\frac{45}{7\alpha^2}}m&=18.64m,\: \mathrm{and} \\[0.3cm]
M_{\calP}&=\sqrt[4]{\frac{45}{\alpha^2}}m&=30.32m.
\end{array}
\eeqn

\subsection{Euler-Heisenberg Dielectric Function}
\label{ssec:QED-eps}
As we will discuss below, the 
kinematical and gravitational effects of a trace contribution to the 
electromagnetic energy-momentum differ strikingly from the classical 
Maxwell energy-momentum \req{classicalTmn}.  
Although this fact should make experimental
verification of the presence of the trace term easy in principle, the 
relative strength of the classical contribution cannot be ignored in 
the study of real physical systems.
Thus, we complete the analysis of the energy-momentum tensor of 
Euler-Heisenberg electromagnetism with evaluation of the dielectric function 
$\veps = -\partial \Leff/\partial\calS$.  

\begin{figure*}
\includegraphics[width=.48\textwidth]{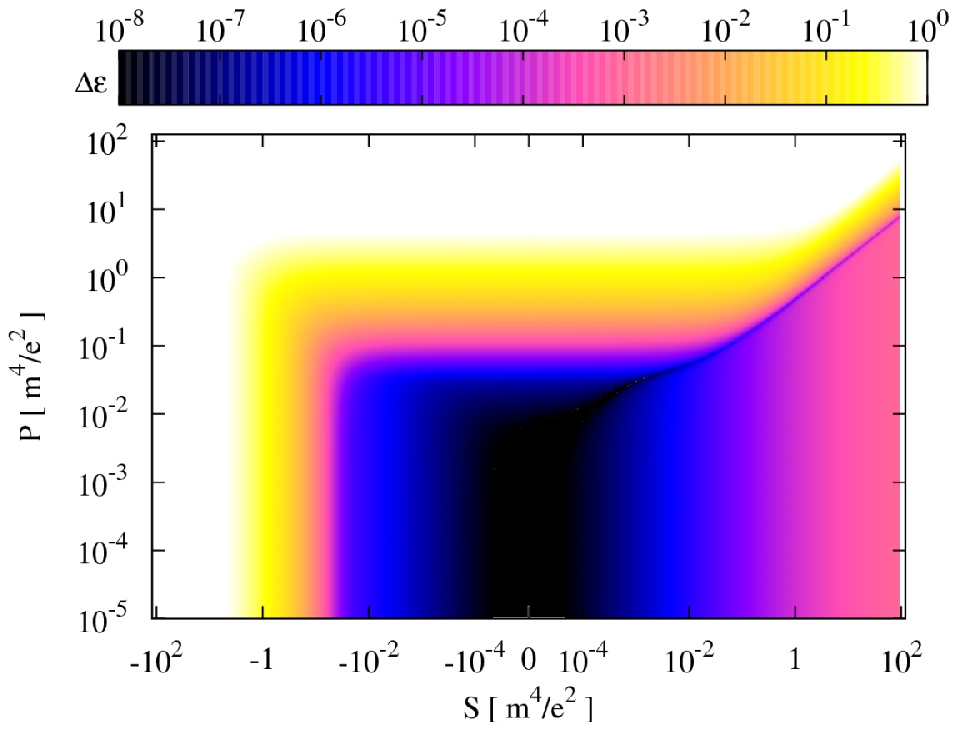}
\includegraphics[width=0.48\textwidth]{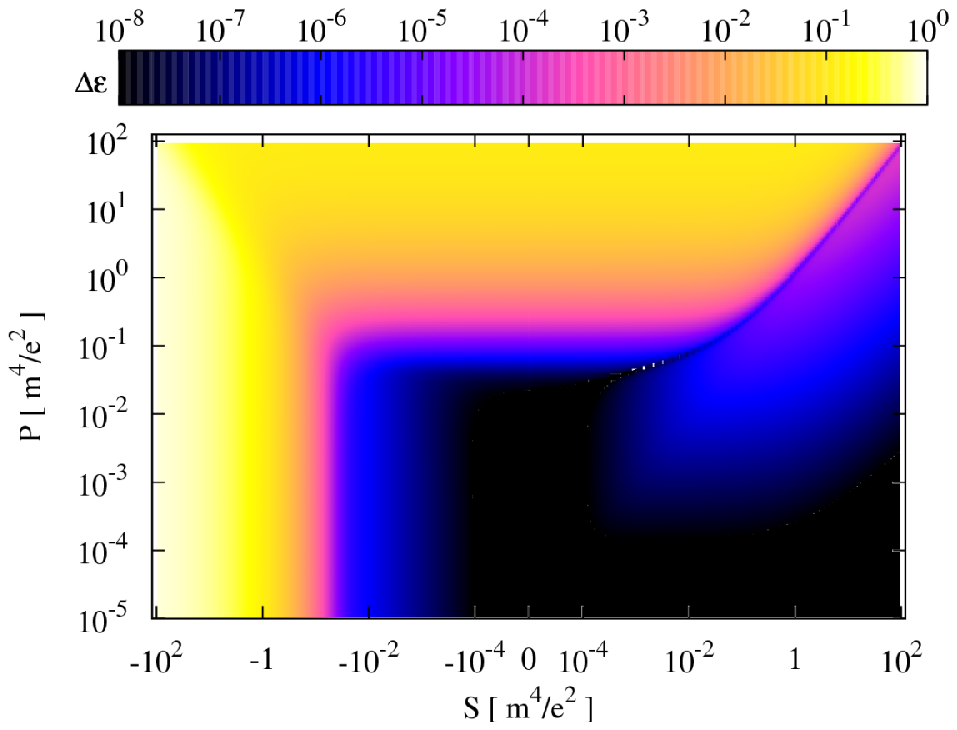}
\caption{The dielectric function for spin-1/2 (left) and spin-0 (right)
quantum fluctuations.  The value is the correction to the
Maxwell $1$, and as in BI theory, a dominantly magnetic field (below
the dark line rising to the right in each plot) gives a 
negative correction suppressing the Maxwell energy-momentum.  
Contours are provided to display larger values obtained at strong
electric fields $(\mathcal{S} < 0)$.}
\label{fig:pol}       
\end{figure*}

The EH actions are corrections to the
classical $-\calS$, so the total dielectric functions have the form
\begin{subequations}\begin{eqnarray}
\veps^{\rm f} -1 &=& - \frac{\partial\Leff^{\rm f}}{\partial\calS}
 \,\defn\,\Delta\veps^{\rm f},\\
\veps^{\rm s} -1 &=& - \frac{\partial\Leff^{\rm s}}{\partial\calS}
 \,\defn\,\Delta\veps^{\rm s},
\end{eqnarray}\end{subequations}
requiring differentiation of the expressions \req{SpinLag} and \req{ScalLag},
via the partial differentials 
$(\partial a/\partial\calS)\partial/\partial a$ and
$(\partial b/\partial\calS)\partial/\partial b$, where
$$ 
\frac{\partial a}{\partial\calS}=\frac{-a}{a^2+b^2}, \quad\mathrm{and}\quad
\frac{\partial b}{\partial\calS}=\frac{b}{a^2+b^2}.
$$

\begin{figure*}
\includegraphics[width=.48\textwidth]{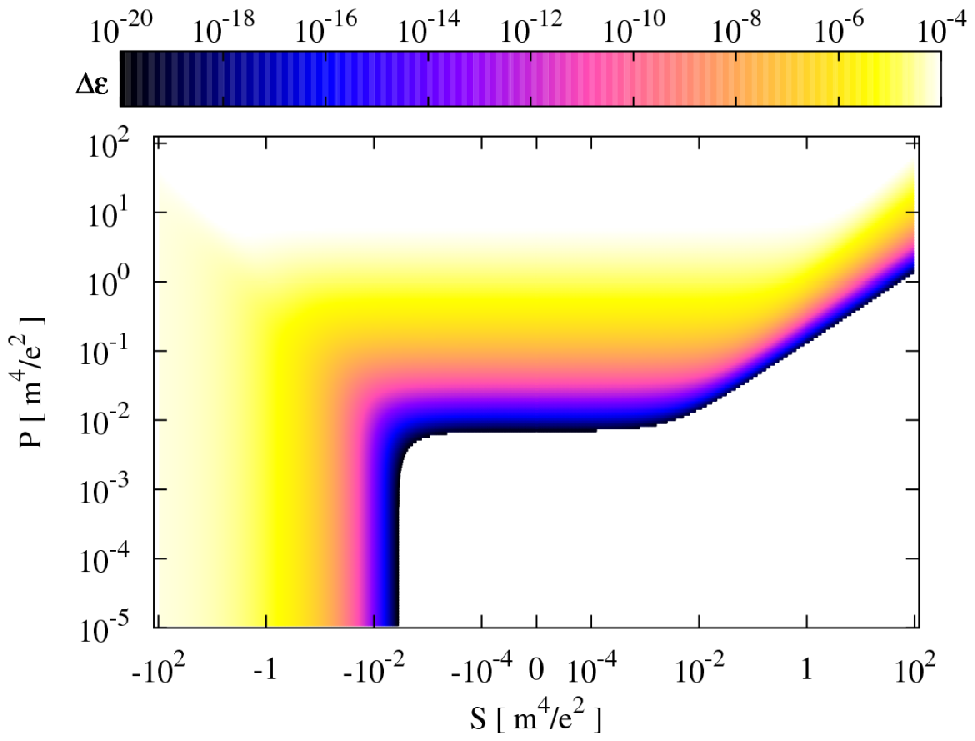}
\includegraphics[width=0.48\textwidth]{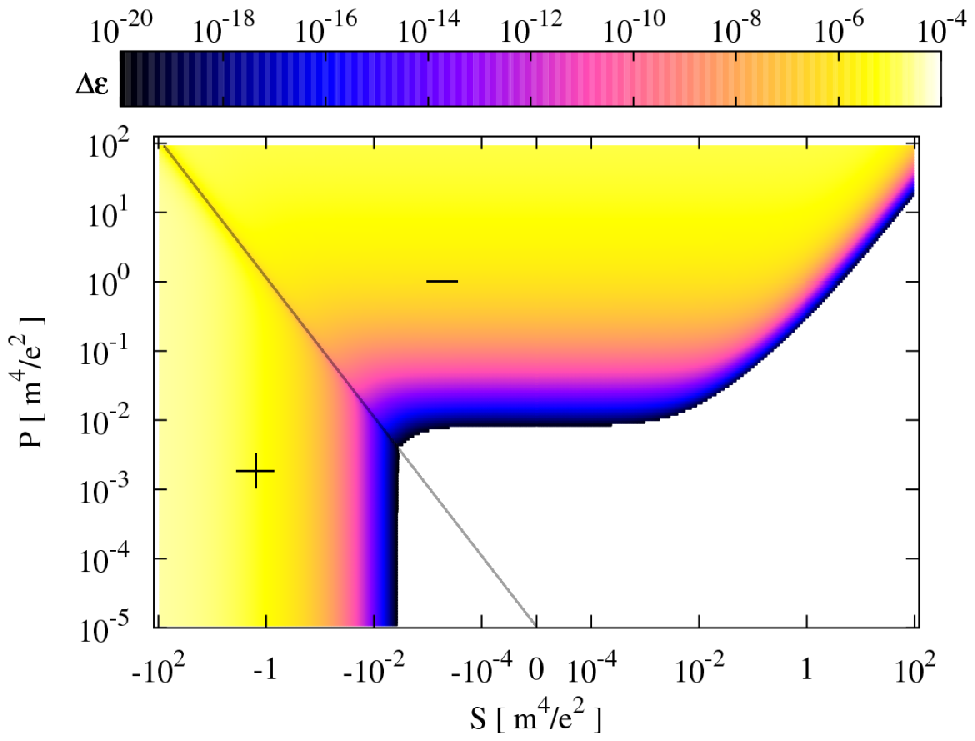}
\caption{The imaginary part of the dielectric function for spin-1/2 (left) 
and spin-0 (right) quantum fluctuations.  For spin-1/2, the dielectric function
is consistently positive; however, as indicated on the plot itself in the 
scalar dielectric function, the imaginary part is positive to the left and 
negative to the right of the overlaid (grey) line at $\mathcal{S}=-\mathcal{P}$.
The vanishing of $\Delta\veps^{\rm s}$ for the anti-self-dual
$\mathcal{S}=-\mathcal{P}$ field configuration is a striking difference from 
the spinor case, which is suppressed when $\mathcal{S} > \mathcal{P}^2$.
Details are suppressed when $\Delta\veps < 10^{-20}$.}
\label{fig:Im-pol}       
\end{figure*}

Considering first the fermionic case \req{SpinLag}, we find
\begin{widetext}
\beqn\label{spin-eps}
\Delta\veps^{\rm f} =
\frac{1}{8\pi^2}\!\int_0^{\infty}\!\! \frac{ds}{s^3}
 \frac{eas\cot(eas)\,ebs\,\coth(ebs)}{a^2+b^2}
\Big(\!
  \coth ebs - ebs\,\csch\!^2ebs - \cot eas + eas\csc^2eas - \frac{2}{3}(b+a)s^2
\Big) e^{-m^2s}\!,
\eeqn
\end{widetext}
for which renormalization only requires subtraction of the logarithmic 
divergence, since the zero-field constant is differentiated away.
Obtaining the meromorphic expansion of the integrand by differentiating
the Sitaramachandrarao identity  \req{Sitarid1}, 
used above in \req{gen-QED-Tmm2}, we have
the numerically more convenient representation
\begin{eqnarray}\label{spin-eps2}
\Delta\veps^{\rm f}\!\!&=&\!\!
\frac{e^4}{2\pi^2}\frac{ab}{a^2+b^2}\int_0^{\infty}\!\! s\:e^{-m^2s}
\left( K_a + K_b \right)ds, \\
K_a \!\!\!&\defn&\!\!\! a^2 \!\sum_{k=1}^{\infty}
 \left(\!\frac{k\pi \coth(k\pi b/a)}{((eas)^2-k^2\pi^2)^2} - 
   \frac{(b/a)\csch\!^2(k\pi b/a)}{(eas)^2-k^2\pi^2}\!\right) \notag \\
K_b \!\!\!&\defn&\!\!\! \!-b^2 \!\sum_{k=1}^{\infty}
 \left(\!\frac{k\pi \coth(k\pi a/b)}{((ebs)^2+k^2\pi^2)^2} + 
   \frac{(a/b)\csch\!^2(k\pi a/b)}{(ebs)^2+k^2\pi^2}\!\right)\!\!. \notag
\end{eqnarray}
This form also provides the imaginary part 
\beqn\label{spin-eps-Im}
\Imag\,\veps^{\rm f} \!=\! 
\frac{\alpha}{2\pi}\frac{\beta\beta'}{\beta^2+\beta'^2}\!\sum_{k=1}^{\infty}\!
\frac{1}{k\pi}\!\left(\!\coth\frac{k\pi\beta}{\beta'}
	-\frac{2\beta}{\beta'}\csch\!^2\frac{k\pi \beta}{\beta'}\!\right)\!
 e^{-k\beta}
\eeqn
which is again a reflection the instability of strong electric fields, 
as confirmed in figure~\ref{fig:Im-pol} by its suppression in dominantly 
magnetic fields.  The polarization function in general field configurations
can also be exhibited in the quasi-statistical form of \req{gen-QED-Tmm3}, 
but the stronger singularity $s^{-3}$ in the proper time variable 
persisting in the absence of the
$m$-differentiation makes for simpler `spectral' functions
$$
\ln(1-s^2+i\eps) \quad\mathrm{and}\quad \ln(1+s^2)
$$
for the electric- and magnetic-like integrals $K_a$ and $K_b$.

The weak field expansion of the dielectric function can be obtained by
straightforward differentiation of the expansion of the effective action, giving
\beqn\label{spin-eps-weakfield}
\Delta\veps^{\rm f} \approxeq
\frac{\alpha}{90\pi}\frac{e^2}{m^4}8\calS
-\frac{2\alpha}{315\pi}\frac{e^4}{m^8}(24\calS^2+13\calP^2)+\ldots
\eeqn
For completeness, we exhibit the dielectric functions for magnetic-only  
\beqn\label{eps-f-Bonly}
\Delta\veps^{\rm f}(B) = 
-\frac{2\alpha}{\pi} \int_0^{\infty} \!\!\!ds\,
 \frac{s(s^2+2)}{(s^2+1)^2}\sum_{k=1}^{\infty}\frac{e^{-k\beta's}}{k^2\pi^2}
\eeqn 
and electric-only 
\beqn\label{eps-f-Eonly}
\Delta\veps^{\rm f}(E) = 
-\frac{2\alpha}{\pi} \int_0^{\infty} \!\!\!ds\,
 \frac{s(s^2-2)}{(s^2-1)^2}\sum_{k=1}^{\infty}\frac{e^{-k\beta s}}{k^2\pi^2}.
\eeqn
backgrounds, recalling $\beta'\to \pi m^2/eB$ and $\beta\to \pi m^2/eE$ 
in the respective limits.

For the scalar case \req{SpinLag},
\begin{eqnarray}\label{scal-eps}
\Delta\veps^{\rm s} \!\!\!&=&\!\!\!
 -\!\int_0^{\infty}\!\frac{e^{-m^2s} \,ds}{16\pi^2s^3}
 \frac{eas\csc(eas)\,ebs\,\csch(ebs)}{a^2+b^2} \\
&& \hspace*{1cm} 
\times \Big( ebs \coth ebs - eas \cot eas - \frac{b+a}{3}s^2 \Big) , \notag
\end{eqnarray}
again renormalized by subtraction of the logarithmic divergence.
The identity used above in \req{gen-scalQED-Tmm}, provides
the numerically more convenient representation
\begin{eqnarray}\label{scal-eps2}
\Delta\veps^{\rm s} \!\!&&\!\!=
 -\frac{e^4}{4\pi^2}\frac{ab}{a^2+b^2}\int_0^{\infty}\!\!ds\, se^{-m^2s}
\left( K_a + K_b \right) \\
K_a \!\defn\!\!&&\!\!\! 
 a^2\!\sum_{k=1}^{\infty} (-1)^k\csch\!\!\!\left(\!\frac{k\pi b}{a}\!\right)
 \!\left(\!\frac{k\pi}{((eas)^2-k^2\pi^2)^2}\right. \notag \\
 && \hspace*{4cm} -
  \left.\frac{(b/a)\coth(k\pi b/a)}{(eas)^2-k^2\pi^2}\!\right) \notag \\
K_b\!\defn\!\!&&\!\!\! 
 -b^2\!\sum_{k=1}^{\infty} (-1)^k\csch\!\!\!\left(\!\frac{k\pi a}{b}\!\right)
 \!\left(\!\frac{k\pi}{((ebs)^2+k^2\pi^2)^2}\right. \notag \\
 && \hspace*{4cm} + 
  \left. \frac{(a/b)\coth(k\pi a/b)}{(ebs)^2+k^2\pi^2} \!\right) \notag
\end{eqnarray}
displaying the imaginary part
\begin{eqnarray}\label{scal-eps-Im}
\!\!\Imag\,\veps^{\rm s} \!&=&\!
-\frac{\alpha}{4\pi}\frac{\beta\beta'}{\beta^2+\beta'^2}
 \!\sum_{k=1}^{\infty}\! \frac{(-1)^k}{k\pi}\,\csch\!\!\left(\!\frac{k\pi \beta}{\beta'}\!\right)\! \\
 && \hspace*{3cm} \times \left(1\! - \!\frac{2\beta}{\beta'}\coth\!\frac{k\pi \beta}{\beta'}\right) e^{-k\beta} \notag
\end{eqnarray}
which is positive only for dominantly electric ($0< \calP < -\calS$) fields.
The limits of magnetic- and electric-only background are obtained 
immediately from Eqs.\,\eqref{eps-f-Bonly} and \eqref{eps-f-Eonly} 
by multiplication with $-1/2$ and insertion of an alternating $(-1)^k$ 
in the sum.  The weakfield expansion is
\beqn\label{scal-eps-weakfield}
\Delta\veps^{\rm s} \approxeq
\frac{\alpha}{1440\pi}\frac{e^2}{m^4}14\calS
 -\frac{\alpha}{20160\pi}\frac{e^4}{m^8}(279\calS^2+77\calP^2)+\ldots
\eeqn

Numerical evaluations of the EH corrections to the spinor and scalar
dielectric functions for general field strengths are displayed in
figure~\ref{fig:pol}.  $\Delta\veps$ again reflects the 
unusually square character of the EH integrals, though in agreement with 
Born-Infeld theory, dominantly magnetic fields suppress $(\Delta\veps<0)$ 
the Maxwell tensor.  However, the boundary for which this magnetic suppression
is present differs between fermionic and scalar electrodynamics, being 
approximately $\calS \propto \calP^2$ in the former case and 
$\calS \propto \calP$ in the latter.  Indeed, the transition to augmentation 
($\Delta\veps > 0$) appears to arise in conjunction with the growing
imaginary part, as seen in figure~\ref{fig:Im-pol}.

\section{Kinematical Effects of $\Tmm$} 
\label{sec:Tmm-kin} 
With numerics providing the magnitude of the induced vacuum 
deformation, we can accurately evaluate physical situations in which the
modification of the Maxwell energy-momentum tensor has observable 
consequences.  
As may be verified by direct calculation (see also discussion in Section 8
of~\cite{BialynickiBirula:1975yp}),
the nonlinear electromagnetism (whether BI or QED) behaves in this regard
as any other nonlinear 
medium and does not alter the Lorentz force
\beqn \label{lorentz-f}
f^{\mu}=j_{\nu}F^{\mu\nu},
\eeqn 
which is dictated by the
necessity of gauge invariance in the coupling of EM potentials to charged 
matter.  This point is to be contrasted with the modification of particle
properties.

Violation of the superposition principle~\cite{JRBI,Dominguez08} entails
an interaction between the background field and the field generated by the 
charged matter.  Such an interaction could be introduced in the Lorentz 
force, but it is more easily evaluated separately using the 
weak-field expansion of the EH effective potential.

As an example, take a large ($r > \lambbar_e$) charged sphere in 
a strong background magnetic field, which could provide a rough
model for an $\alpha$-particle in the atmosphere of a highly magnetized
neutron star.
In the rest frame of a non-relativistic charged probe particle, we take the 
background magnetic field as constant $\vec B = B\hat z$ and the electric 
field as the particle's Coulomb  field $\vec E = Ze\hat r/r^2$.  
Integrating the energy of the combined field configuration, $T^{00}$,
over the volume with
a short distance cutoff at the Compton wavelength $\lambbar$,
the leading contribution is the trace  $\Tmm \to u_{\rm eff}$:
\beqn\label{EH-NLO}
u_{\rm eff}
 = \!\int \!d^4x\,\frac{2\alpha^2}{45m_e^4}(7\mathcal{P}^2\!+4\mathcal{S}^2)
 = \frac{2\alpha^2}{45m_e^4}\frac{4\pi}{3\lambbar}(ZeB)^2,
\eeqn
keeping only the nonlinear-sourced cross terms.  The coefficient of the
Maxwell energy $-\frac{\partial\Leff}{\partial\mathcal{S}}$ also induces 
cross terms subleading at $\mathcal{O}(\alpha^3)$.  The cutoff arises since 
at distances shorter than $\lambbar$ we must use quantum dynamics to 
describe the probe particle, consideration of which would be inconsistent 
with the classical particle dynamics.

This interaction energy is positive, independent of the sign of the charge, 
and comparable to the gravitational potential of a neutron star with 
dipolar magnetic field.  As $u_{\rm grav}$ 
is negative and $\propto r^{-1}$ and the effective (scalar) potential goes 
with $B^2\propto r^{-6}$, 
\beqn\label{pot-ratio}
\frac{u_{\rm eff}}{u_{\rm grav}} =
\frac{8\pi(\alpha Z)^2}{135\lambbar}\!
 \left(\!\frac{eB_{\rm surf}}{m_e^2}\!\right)^{\!2}\!\frac{R^6_{\rm surf}}{r^6} 
  \left(\! \frac{1.48 M_{\odot}m}{r}\!\right)^{\!-1}
\eeqn
converting Newton's constant into the convenient units $G=1.48$\,km/solar mass. 
The $\lambbar$ cutoff in Eq.\,\eqref{EH-NLO} cancels against particle mass $m$,
making Eq.\,\eqref{pot-ratio} independent of both the mass of the particle $m$ 
and the cutoff $\lambbar$.
Remarkably, at the surface of a 1.5 $M_{\odot}$, 14 km radius star with
critical surface field $B_{\rm surf} = B_c$, the nonlinear-electromagnetic
effective potential is 34 times the gravitational potential, resulting in  
a large repulsive, quasi-Lorentz-scalar potential  for charged particles 
entering the strong field region. 

For a relativistic particle, the stellar magnetic field is Lorentz-transformed
requiring consideration of further cross-terms.  The calculation is 
simplified by choosing a frame, though the effective potential is still 
determined from \req{EH-NLO}.  $\gamma$-factors from the Lorentz 
transformation enhance the effect, but terms linear in the field
of the particle introduce charge dependence.

\section{Discussion and Conclusions}\label{SecDC}

The central motivation of this study is the observation that 
externally applied fields in nonlinear electromagnetism have a 
dark energy-like contribution to their energy-momentum tensor.
We therefore examined the physics giving rise to the trace
of the energy-momentum tensor as an avenue of insight into the 
origin of the observed dark energy in the universe.  As $T^{\mu}_{\mu}$
in our study is generated by quantum-induced nonlinearity of the
electromagnetic field the physics of dark energy is accessible 
to laboratory experiment probing electromagnetism at high fields.

We derived the energy-momentum tensor for general nonlinear electromagnetic
theories and emphasized the form \req{SchTmn1}.
We considered the relationship of the trace with the 
matter condensate and obtained a result, Eq.\;(\ref{QED-Tmm}) 
which amounts to removal of the leading 
term in $\calS$ in \cond. This reduces the numerical results by 
a factor of about 100 and along with this, the
physical dark energy effect of the energy-momentum trace is greatly reduced. 

Employing  the resummation technique introduced in~\cite{Mueller77}, 
we numerically evaluated the deviations from the Maxwell tensor, the 
dielectric function and the trace for Born-Infeld electromagnetism and 
the Euler-Heisenberg effective action with both Fermi and Bose matter 
fields. We believe that these are the first presentations in literature 
of the matter condensate and energy-momentum trace 
arising from the Euler-Heisenberg action at field strengths well beyond 
critical.  The dielectric function is found in both BI and EH to suppress 
the Maxwell tensor in the presence of dominantly magnetic fields.  
The dielectric response of the vacuum thereby enhances the observable
consequences of the presence of the energy-momentum trace.

The Born-Infeld theory is at first sight an interesting source of 
energy-momentum trace.  However, a lower limit on the
limiting BI electric field strength obtained 30 years 
ago from the study of precision atomic and muonic
spectra~\cite{JRBI} requires $M^2/e\ge 1.7 10^{22}$V/m, 
implying $M\ge 60$MeV.  Contemporary $g-2$ experimental
results, if analyzed with the objective to set a limit 
on the BI scale, would very probably push this limit further up.
As $g-2$ of the electron in strong fields is in itself a project 
requiring study of the two-loop Lagrangian~\cite{Ritus}, such analysis
takes us far beyond the scope of this paper, and we leave the investigation
to the future.  The scale of the energy-momentum trace of BI type must be 
rather large and the effect at best comparable to the effect
of vacuum fluctuations expressed by the Euler-Heisenberg effective action.
Even though the energy-momentum trace is suppressed by the QED 
coupling, $(\alpha/\pi)^2=6\E{-6} $ (see section~\ref{sec:QED}), 
the effective scale $40\,m_e\simeq 20$ MeV implies that quantum fluctuations
remain dominant compared with any other current theoretical 
framework, given the experimental  constraints. 

This is consistent with BI being a high-cutoff theory arising from 
more fundamental matter properties.  Hence, we devoted the largest part of 
this report to expanding our prior study of electron fluctuations in the 
vacuum, the energy-momentum trace has previously been discussed
as a signal of vacuum deformation~\cite{Labun:2008gm}.
The energy density in the trace is then interpreted as the 
shift of the vacuum energy induced by the applied field, and
in concordance with this interpretation, the trace is positive
definite when extracted correctly from the effective action.  The smooth BI
$\Tmm$ highlights the extraordinary form of the quantum-induced trace. 
Prior considerations of the analytic structure of the Euler-Heisenberg 
effective action had not made apparent the near singular boundaries present  
in the condensate and the trace.  The present evaluations suggest further
investigations into the strong field vacuum phase structure.

We provided further the first complete numerical calculations
of the electron-positron condensates of spinor and scalar QED for arbitrary
fields and have addressed the contradictory claims in literature 
relating to claims that $\Tmm$ and $-m\cond$ are equal, and we found in 
our non-perturbative study a clear difference originating in the scale 
dependence of charge renormalization.
We emphasize that our evaluation of the condensate and of the 
energy-momentum trace are completely independent of renormalization procedure.

Being a vacuum phenomenon, the QED $\Tmm$ studied here is a volumetric, 
extrinsic property, so to obtain observable effects in the laboratory, one 
could imagine creating a macroscopic region of ultra-strong
electromagnetic field.   On the other hand, as the energy-momentum trace 
of QED is induced by external fields, the effects exist anywhere and everywhere
an electromagnetic field is present.  For instance, one computes that to 
generate the observed dark energy density
\begin{equation}\label{dark}
\frac{\Lambda}{4\pi G}\simeq (2.325{\rm meV})^4 
	\simeq 6.09\E{-10} {\rm J/m}^3
\end{equation} 
corresponds to the trace induced by a magnetic field of $108$\,T spanning
the universe, clearly an unphysical condition.  The energy-momentum trace
induced by nonlinear electromagnetism indicates the possibility of observing
anti-gravity-like effects in the laboratory with high intensity laser 
experiments.

Although many previous studies of vacuum energy and issues 
arising from conformal symmetry breaking have focused on QCD,
we did not address here how the structure of 
the QCD vacuum, which is very strongly deformed by 
glue and quark fluctuations, relates to the trace \Tmm 
and responds to an applied electromagnetic field.  
Some discussion of this question, including its relation to dark 
energy, has been already offered~\cite{Schutzhold:2002pr}.
Combining the quantum vacuum with general relativity remains today a very 
delicate question however~\cite{Weinberg:1988cp}.   
Our report establishes an important connection, tying already-recognized
quantum vacuum effects to  dark energy.

To summarize, we have studied the energy-momentum tensor of nonlinear 
electrodynamics emphasizing an explicit relationship of the dark energy-like
trace of the energy-momentum tensor \req{Tan} to the 
nonlinearity of the theory.  In the consideration of electrodynamics
as a quantum gauge theory, the connection provided a new derivation of a
non-perturbative identity between the energy-momentum trace and the gauge 
and matter condensates, \req{QED-Tmm}.  The 
Euler-Heisenberg effective action provided a natural example for the 
numerical evaluation of the condensate and energy-momentum trace, the results 
of which are displayed for both fermionic and scalar fields in  
figures~\ref{fig:gen-Tmm} and~\ref{fig:scal-genTmm}.  Finally, we briefly 
explored the implications of an energy-momentum trace for charged-particle
kinematics.\\

\acknowledgments
We thank Prof. D. Habs, Director of  the Cluster 
of Excellence in Laser Physics --  Munich-Center for Advanced Photonics (MAP)
for   hospitality in Garching where this research was in part carried out.
This work was supported by the DFG Cluster of Excellence MAP 
(Munich Centre of Advanced Photonics), and by a grant
from: the U.S. Department of Energy  DE-FG02-04ER41318. 

\begin{appendix}
\section{Improving Convergence of Euler-Heisenberg Integrals}\label{app:stats}
In this appendix, we display the steps in the transformation of the proper time 
integrals Eqs.\,\eqref{SpinLag} and \eqref{ScalLag} into the more rapidly
convergent representations used for numerics in the text.

The effective action for electrodynamics displays non-analyticities
that, generating an imaginary part of the action, are associated with the
instability of the vacuum.  However, our method of resumming the poles is 
very useful for improving the overall convergence of integrals of the 
Euler-Heisenberg form, and we start with the case of only a magnetic
field being present, for which 
\beqn\label{spin-lag-Bonly}
-m\cond = \frac{m^2}{4\pi^2}\int_0^{\infty}\frac{ds}{s^2}(eBs \coth eBs -1)
\eeqn
is analytic on the real axis.  We use the (subtracted) meromorphic expansions, 
\begin{eqnarray} 
x \coth x - 1 
 &=& 2x^2\sum_{k=1}^{\infty}\frac{1}{x^2+k^2\pi^2} \label{Bmerom}\\
 &=&\frac{x^2}{3}-2x^4\sum_{k=1}^{\infty} 
             \frac{1}{(k\pi)^2}\frac{1}{x^2+k^2\pi^2}. \label{BmeromSub}
\end{eqnarray}
Inserting \req{Bmerom} in \req{spin-lag-Bonly} we obtain
\beqn
-m \cond = \frac{m^2 (eB)^2}{2\pi^2}\int_0^{\infty}ds
            \sum_{k=1}^{\infty}\frac{e^{-m^2s}}{(eBs)^2+(k\pi)^2}.
\eeqn
All terms are individually absolutely convergent, so we reorder the sum and
integral following the procedure in~\cite{Mueller77}.  After rescaling 
$s\to s k\pi /eB $, the $k$-sum is evaluated in closed form and we obtain
\beqn\label{app:condB}
-m\cond = -\frac{m^4}{2\pi^2\beta'}\int_0^{\infty}
		\frac{\ln(1-e^{-\beta's})}{1+s^2}\,ds,
\eeqn
which is \req{statkern-B}.

\req{BmeromSub} allows us to remove the quadratic term in \req{QED-Tmm-int}, 
and, rescaling and resumming, we find \req{Tmm-B-step2}.
A further integration by parts results in
\beqn\label{app:TmmB} 
\Tmm^{\rm f} = \frac{m^4}{2\pi^2}\int_0^{\infty}
	\frac{s-\mathrm{Arctan}\,s}{e^{\beta's}-1}\, ds
\eeqn
which is the $a \to 0$ limit of \req{gen-QED-Tmm3}, but we retain the form
\req{Tmm-B-step2} for clarity in the associated discussion of signs.

Turning now to $B=0$, i.e. electric field only with $a \to |E|$, we see in 
the meromorphic expansion  
$$x\cot x-1=2x^2\sum_{k=1}^{\infty}\frac{1}{x^2-k^2\pi^2}$$
the singularities that indicate the instability of the system to produce 
real pairs.  We assign to the mass a small imaginary component 
$m^2\to m^2+i\epsilon$ so that
\beqn \label{cond-E-step1}
-m\cond =\frac{m^2(eE)^2}{2\pi^2}\int_0^{\infty}\!\!ds
   \sum_{k=1}^{\infty}\frac{e^{-m^2s}}{(eEs)^2-(k\pi)^2+i\eps},
\eeqn
whence resummation produces \req{statkern-E}.
Removing the leading term in meromorphic expansion for the case of 
the electric field by use of 
$$
x\cot x-1 =
-\frac{x^2}{3}+2x^4\sum_{k=1}^{\infty}\frac{1}{k^2\pi^2}\frac{1}{x^2-k^2\pi^2},
$$
we obtain \req{Tmm-spinorE}.

For general fields, the proper time integrals are rewritten
using the Sitaramachandrarao identity (Eq.\,(6) in ~\cite{Cho00})
\begin{eqnarray}\label{Sitarid1}
xy \coth x \cot y \!\!&=& \!\! \notag
1\!+\!\frac{x^2-y^2}{3}
-2x^3y\!\sum_{k=1}^{\infty}\frac{1}{k\pi}\frac{\coth(k\pi y/x)}{x^2+k^2\pi^2} \\
&&\hspace*{1cm} 
+\,2y^3x\!\sum_{k=1}^{\infty}\frac{1}{k\pi}\frac{\coth(k\pi x/y)}{y^2-k^2\pi^2} 
\end{eqnarray}
with the result
\begin{subequations}\label{gen-QED-Tmm}
\begin{eqnarray} 
&&~\Tmm^{(QED)}=\frac{m^4}{2\pi^2} \left( I_a + I_b \right); \\
I_a \!&=&\! e^4a^3b\!\int_0^{\infty}\!\!\!ds s^2e^{-s}\sum_{k=1}^{\infty}
       \frac{\coth(k\pi b/a)}{k\pi( k^2\pi^2-(eas)^2)}, \\
I_b \!&=&\! e^4b^3a\!\int_0^{\infty}\!\!\!ds s^2e^{-s}\sum_{k=1}^{\infty}
        \frac{\coth(k\pi a/b)}{k\pi( k^2\pi^2+(ebs)^2 )}, 
\end{eqnarray}
\end{subequations} 
Rescaling converts these expressions to those found in \req{gen-QED-Tmm2}.

The quasi-statistical representation of $I_a$ in \req{gen-QED-Tmm3} is
derived by exchanging the sum and the integral in order to integrate by parts,
\beqn\label{Ia-step1}
I_a \!=\! -b\!\int_0^{\infty}\!\!\!\!ds
(s+\frac{1}{2}\ln\!\left(\!\frac{1-x+i\eps}{1+x-i\eps}\!\right) )
\sum_{k=1}^{\infty}\!\coth\!\left(\!\frac{k\pi a}{b}\!\right)\!
e^{-\frac{k\pi}{ea} m^2s}
\eeqn
We break up the $\coth$ function:
\beqn
\coth x = \sum_{\sigma=\pm 1} \frac{e^{\sigma x}}{1-e^{-2x}}e^{-x}.
\eeqn
Expanding the denominator as a power series, the sum in \req{Ia-step1}
becomes
\beqn\label{knsigma-sum}
\sum_{k,n,\sigma} \exp\left( -\frac{k\pi}{ea}(m^2 s + (2n+1-\sigma)eb)\right).
\eeqn
Since $\exp(-k\pi b/a)<1$, the $n$-sum is absolutely convergent, though slowly
when $b/a \ll 1$.  We can exchange the order of summation and do the $k$-sum:
\beqn\label{Ia-step2}
I_a = -b\int_0^{\infty}\!\!\!ds
(s+\frac{1}{2}\ln\left(\frac{1-x+i\eps}{1+x-i\eps}\right))
\sum_{n,\sigma}\frac{e^{-\beta\mathcal{H}_{\sigma}}}{1-e^{-\beta\mathcal{H}_{\sigma}}}
\eeqn
with $\beta \equiv \pi m^2/ea$ and
$m^2\mathcal{H}_{\sigma}$ the inner expression in \req{knsigma-sum},
thus obtaining \req{gen-QED-Tmm3}.

For the scalar case, \req{ScalLag}, 
we require identities paralleling those 
used in the spinor integrations:
\begin{subequations}\begin{eqnarray}
x\:\mathrm{csch}\:x -1\!\!=&&\!\!\!
    -\frac{1}{6}x^2\!-2x^4\!\sum_{k=1}^{\infty}\! \frac{1}{(k\pi)^2}\frac{(-1)^k}{x^2+k^2\pi^2}\\
 \! \!=&&\!  2x^2\sum_{k=1}^{\infty} \frac{(-1)^k}{x^2+k^2\pi^2}
\end{eqnarray} \end{subequations}
\begin{subequations}\begin{eqnarray}
x \csc x -1\!\!\!=&&\!\!\frac{1}{6}x^2+2x^4\!\sum_{k=1}^{\infty} \frac{1}{(k\pi)^2}\frac{(-1)^k}{x^2-k^2\pi^2}\\
\!\!\!  =&& \!\!2x^2\sum_{k=1}^{\infty} \frac{(-1)^k}{x^2-k^2\pi^2}
\end{eqnarray}\end{subequations}
The representation for general fields uses
an identity closely related to \req{Sitarid1} 
(see Eq.\:(20) of~\cite{Cho00}):
\begin{eqnarray}\label{Sitarid2}
xy\:\csch x \csc y \!\!&=& \!\!\!
1\!+\!\frac{x^2-y^2}{6}
-2x^3y\!\sum_{k=1}^{\infty}\!\frac{(-1)^k}{k\pi}\frac{\csch\!(k\pi y/x)}{x^2+k^2\pi^2} \notag \\
&&\hspace*{0.8cm} 
+\,2xy^3\!\sum_{k=1}^{\infty}\!\frac{(-1)^k}{k\pi}\frac{\csch\!(k\pi x/y)}{y^2-k^2\pi^2}
\end{eqnarray}

For the further resummation resulting in the statistical representation, 
the csch function is expanded analogously 
\beqn
\csch x = \frac{2}{1-e^{-2x}}e^{-x} = e^{-x}\sum_{n=0}^{\infty}e^{-2nx}
\eeqn
but the absence of cosh in the numerator means no sum over $\sigma$
is introduced.  The remaining procedure is the same.
\end{appendix}

\vfill


\end{document}